\documentclass[a4paper, 10pt, conference]{ieeeconf}      

\IEEEoverridecommandlockouts                              

\usepackage{graphics} \usepackage{epsfig} \usepackage{float} 

\usepackage{mathptmx} \usepackage{times} \usepackage{amsmath} \usepackage{amssymb}  

\usepackage{dsfont}
\usepackage{soul}
\usepackage[normalem]{ulem}
\usepackage[utf8]{inputenc}
\usepackage[small]{caption}
\usepackage{graphicx}
\usepackage{amsmath}
\usepackage{booktabs}
\usepackage{algorithm}
\usepackage[noend]{algorithmic}

\usepackage{amsfonts}
\usepackage{blindtext}
\usepackage{xspace}
\usepackage{subfig}
\usepackage{multicol, blindtext}
\graphicspath{{Figures//}{Figures/All_Figures//}{Figures/All_Figures/Graphs//}{Figures_final_paper_submission//}{img//}}
\DeclareGraphicsExtensions{.eps,.pdf,.png,.jpg,.tif,.tiff,.ps}

\usepackage{caption}
\usepackage{tabu}

\usepackage[belowskip=-15pt,aboveskip=5pt]{caption}
\usepackage[export]{adjustbox}

\makeatletter
\let\NAT@parse\undefined
\makeatother
\usepackage[hidelinks]{hyperref}  \usepackage{url}
\usepackage[nameinlink,capitalize]{cleveref}
\usepackage{soul}
\usepackage{xcolor}
\definecolor{navy}{rgb}{0.1, 0.1, 0.8}
\definecolor{gray}{rgb}{0.6, 0.6, 0.6}
\definecolor{myblue}{rgb}{.8, .8, 1}
\definecolor{olive}{rgb}{0.1, 0.5, 0.1}
\newcommand{\eat}[1]{}

\usepackage[colorinlistoftodos, textwidth=10mm]{todonotes}

\usepackage[numbers]{natbib}

\newcommand{\titlename}{Graph modelling approaches for motorway traffic flow prediction}

\title{\LARGE \bf \titlename}

\author{Adriana-Simona Mihaita$^{1}$, Zac Papachatgis$^{1}$ and Marian-Andrei Rizoiu$^{1}$
\thanks{$^{1}$Adriana-Simona Mihaita, Zac Papachatgis and Marian-Andrei Rizoiu are with the UTS Data Science Institute, University of Technology in Sydney,
81 Broadway Ultimo, NSW, Australia. 
Corresponding author's contact: {\tt\small \url{adriana-simona.mihaita@uts.edu.au}.}}
}

\begin{document}

\maketitle
\thispagestyle{empty}
\pagestyle{empty}

\begin{abstract}
Traffic flow prediction, particularly in areas that experience highly dynamic flows such as motorways, is a major issue faced in traffic management. Due to increasingly large volumes of data being generated every minute, deep learning methods have been used extensively in the latest years for both short and long term traffic flow prediction. However, such models, despite their efficiency, need large amounts of historical information to be provided, and they take a considerable amount of time and computing resources to train, validate and test. This paper presents two new spatial-temporal approaches for building accurate short-term predictions along a popular motorway in Sydney Australia, by making use of the graph structure of the motorway network (including exits and entries). Our proposed methods are proximity-based, and they use the most recent available traffic flow information of the upstream counting stations closest to a given target station. 
Where such information is not available they employ daily historical means instead.
We show that for short-term predictions (less than 10 minutes into the future), our proposed graph-based approaches outperform state-of-the-art deep learning models, such as long-term short memory, convolutional neuronal networks or hybrid models.
\end{abstract}

\section{INTRODUCTION}\label{S1_Intro}

\nocite{appendix}

Traffic flow prediction is an essential element of any Intelligent Transportation System, as the traffic flow data is a key element for accurately determining both seasonal factors (such as the times of congestion, seasonality of traffic flows, impact of public holidays), as well as stochastic factors (abnormal flows due to events, impact of weather, etc). 
Particularly for the stochastic factors, accurate and timely forecasts are paramount to understanding these issues, so they can be actioned, before or as they arise. 
However, obtaining accurate prediction results still represents a challenge due to several reasons such as: 
a) the large amounts of data sets generated every minute across large areas, 
b) the spatial structure and layout of the network can induce high complexity in the localisation of traffic count stations and their utilisation, 
c) the stochastic events which can severely disturb regular traffic conditions, 
d) the spatial and temporal distribution of traffic flow can induce direct and indirect congestion propagation patterns and 
e) missing or erroneous data due to varying equipment functioning state, or inconsistent human reporting. 
All these factors make the prediction exercise a very challenging task. 
 
\textbf{Deep Learning for traffic flow prediction.}
Across the years, there have been several models and prediction approaches being developed.
Auto-regressive models, starting with ARIMA in the early 1990s~\citep{VANDERVOORT1996307}, made use of traffic flow's strong seasonality. 
Such models were further extended by SARIMA \citep{Khoei2013} and ARIMAX \citep{Williams2001}. 
Auto-regressive models were later surpassed by the Deep Learning (DL) model BPNN \citep{DOUGHERTY199721} in 1997, which was shown to produce lower error scores overall, marking the beginning of successful application of deep learning approaches for traffic flow prediction.
From 2015 onwards, the number of proposed approaches in the area of traffic flow prediction began to rise. 
Stacked Autoencoders (SAE) were introduced in 2015~\citep{Lv2015}, followed by the widely referenced convolutional neuronal network CNN in 2016~\citep{Zhang2016}. 
In the same year DLL, DLM8L, DLM8, DLTF15, VARM8L and VARTF15L methods were introduced by \citet{POLSON20171}, who found that ``future traffic conditions are more similar to current ones as compared to those from previous days''. 
A second leap in using Deep Learning to traffic flow prediction came with the usage of Recurrent Neural Networks (RNN), which were designed for sequence modelling and had already revolutionised fields such as Natural Language Processing and automatic translation.
Such a model is
the Short-Term Memory Model (LSTM) \citep{Dai2017} popularised in 2017. 
Its extensions were designed to accommodate external factors such as rain, and led to models such as R-LSTM, R-DPN, R-BPNN and R-ARIMA \citep{Jia2017}. 
Starting from 2018, a series of hybrid deep-learning models have been proposed~\citep{WU2018166},
generating debates on whether hybrid modelling brings a true benefit when accounting for the increase in architecture complexity, and sometimes even a decrease in the prediction accuracy \citep{Mihaita2019}.  

\begin{figure*}[tbp]
    \centering
	\includegraphics[width=1\textwidth]{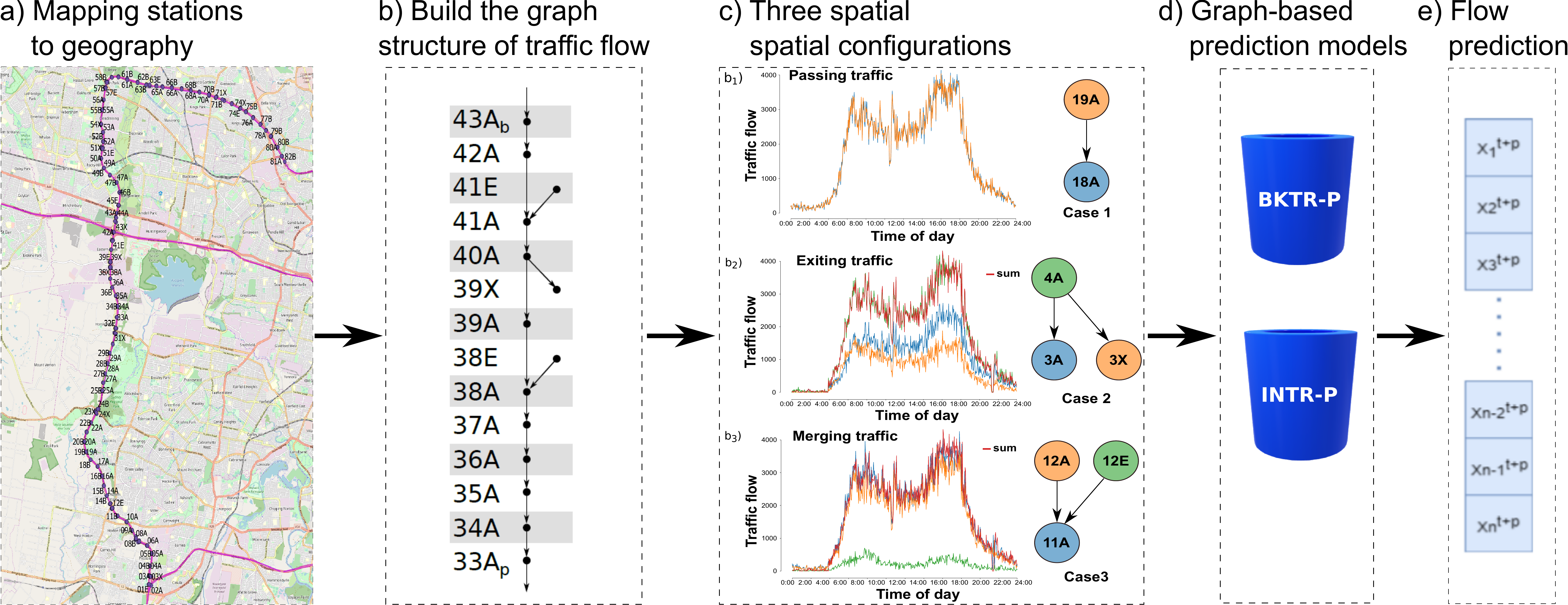}
	\caption{
		Methodology for the proposed graph-flow prediction along the M7 motorway, as a succession of five steps.
	}
\label{Fig2_1_Methodology}
\end{figure*}

\textbf{Graph modelling approaches.}
Another family of approaches gaining popularity aim to
integrate the spatial structure of the transport network into the prediction model -- dubbed \emph{graph modelling approaches}.
Various such methods have been proposed, such as graph-convolution neuronal networks (e.g. GRNN~\citep{Wang2018}, STGCN~\citep{Yu2017}, DCRNN~\citep{Wang2018}) or spatio-temporal approaches such as GTSNet~\citep{Shen2019}.
These approaches provide boosts of performance, however suffer the typical complexity overhead of Deep Learning tools.

\textbf{The work in this paper}
proposes two approaches -- dubbed \emph{backtracking} (BKTR-P)  and \emph{interpolation} (INTR-P) -- that leverage the structure of the motorway network to predict traffic flow.
Our graph modelling approaches join the spatial and temporal traffic flow information in a dynamic fashion, without incurring the complexity of deep learning.
To predict traffic for a given station, both methods rely on the recent traffic flow information received from closest interconnected traffic flow stations, which are aggregated to reflect the traffic flow dynamics on a real motorway network from Sydney, Australia. 
The prediction performances are compared against state-of-the-art deep learning approaches that we have deployed for the network (including hybrid modelling of CNN-LTSM).
We showcase an outstanding accuracy for short-term traffic prediction.

\section{Methodology}\label{S2_Method}
In this section we introduce our proposed graph-based methodology for predicting the traffic flow along motorways. 
\cref{Fig2_1_Methodology} summarises the methodological framework, which consists of five steps: 
a) mapping stations to the motorway geography, 
b) building the graph of traffic flow between stations, 
c) analysing three spatial configurations of stations and flow (all detailed in \cref{subsec:setup-notations}), 
d) proposing two graph-based prediction models presented in order of their efficiency (detailed in \cref{subsec:historical-data,subsec:models}) and 
finally, obtaining e) the traffic flow prediction results.

\begin{figure*}[tbp]
	\centering
	\newcommand\myheight{0.197} 

	\subfloat[]{
		\includegraphics[height = \myheight\textheight]{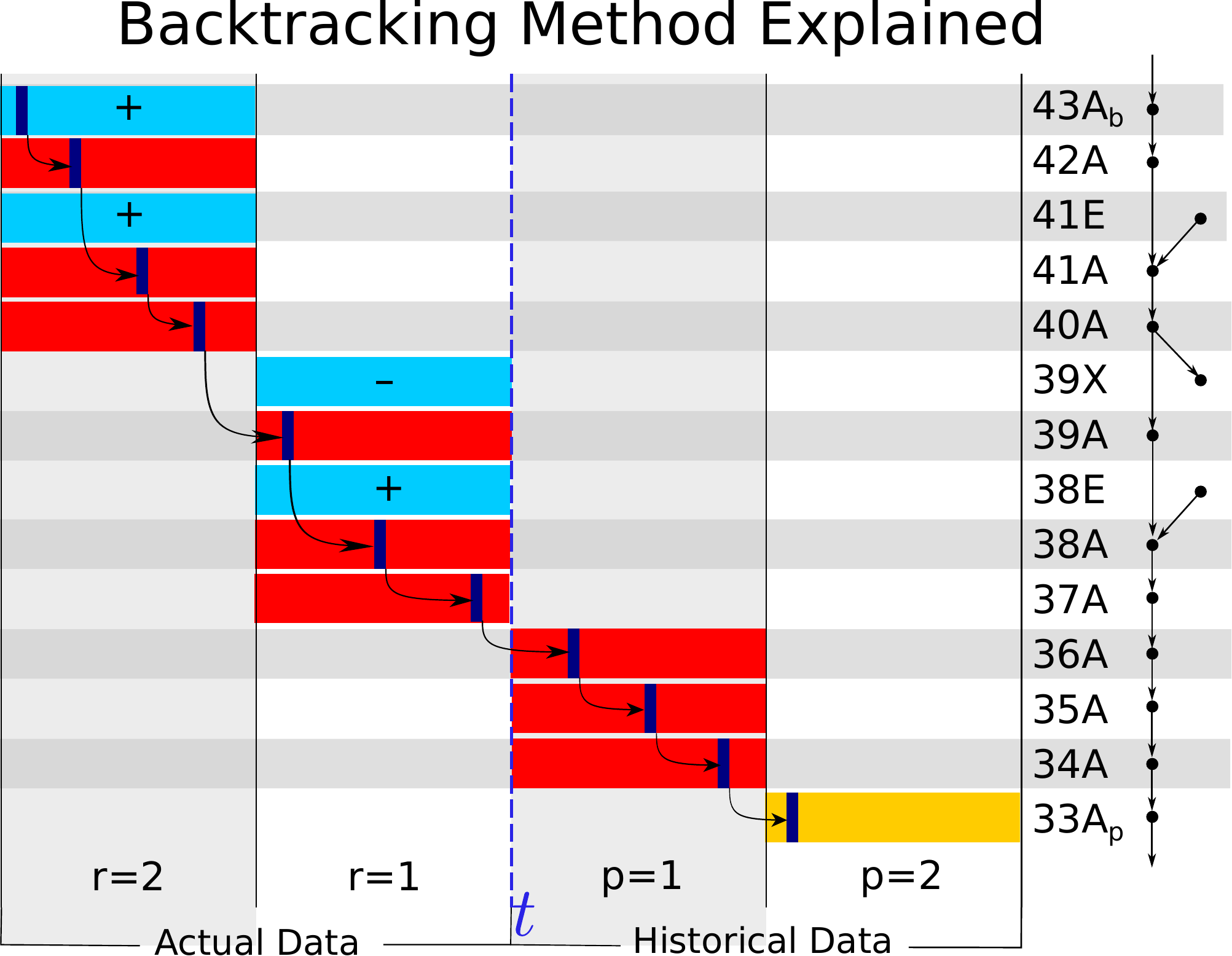}\label{subfig:backtrack-schema}}
	\subfloat[]{
	    \raisebox{0.35\height}{\includegraphics[height=0.12\textheight]{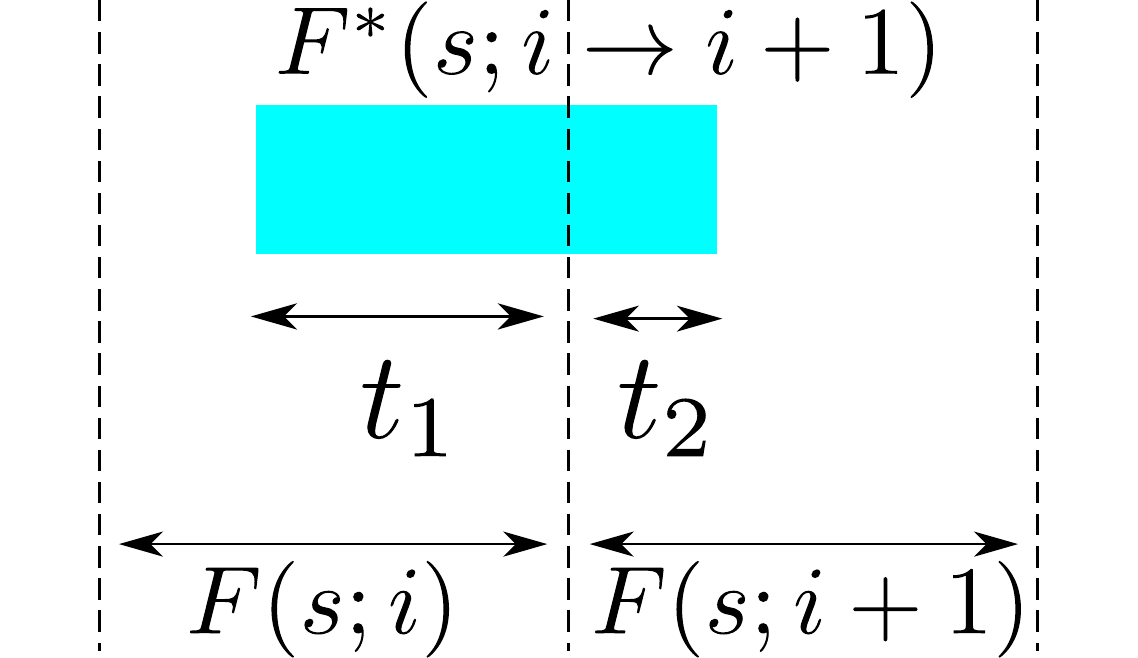}}\label{subfig:interpolation-explain}}
	\subfloat[]{
		\includegraphics[height = \myheight\textheight]{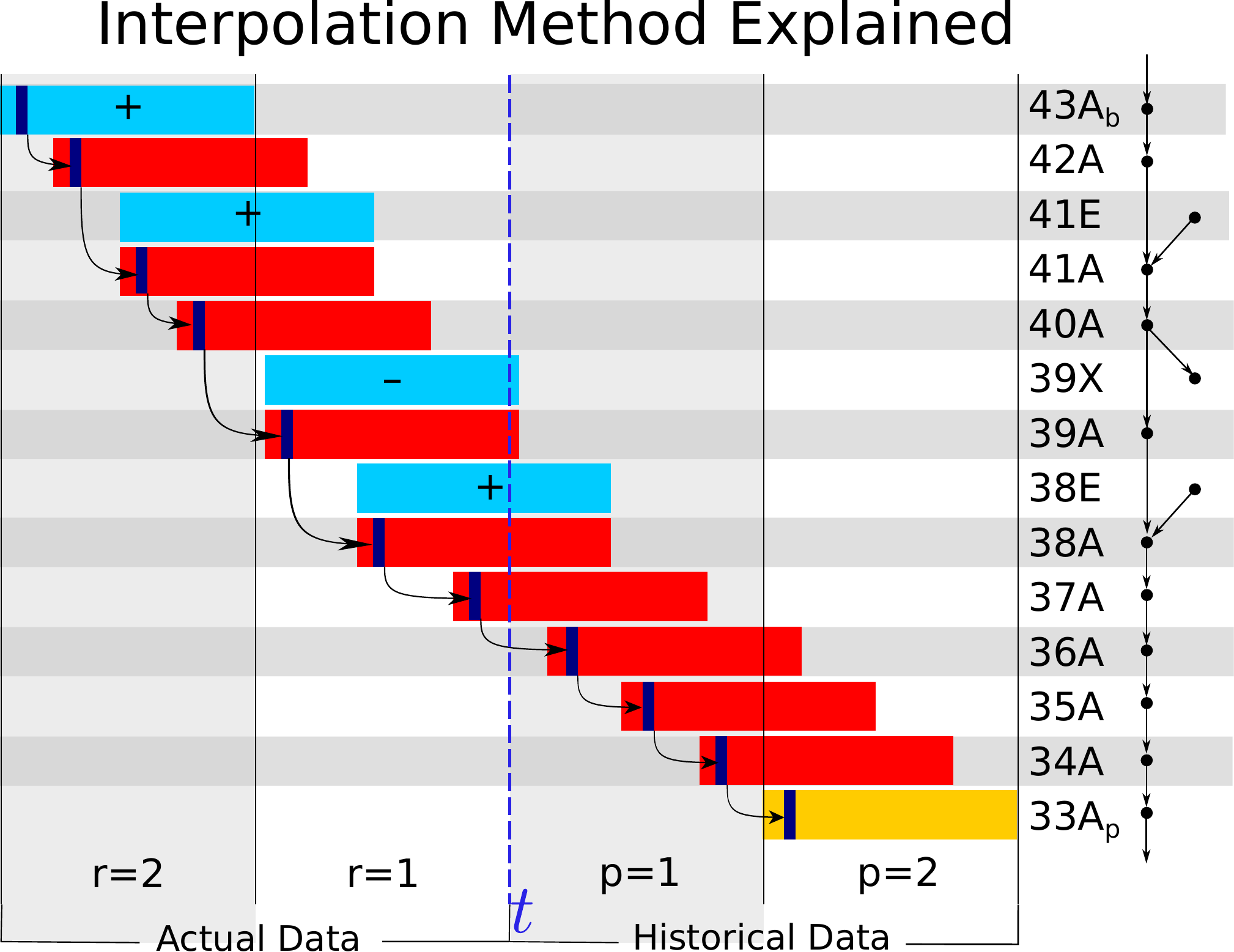}
		\label{subfig:interpolation-schema}
	}
	\caption{ Graphical representation for Backtracking and Interpolation methods on a selected motorway portion: a) BKTR-P exemplification, b) flow representation between two time intervals and c) INTR-P exemplification.}
	\label{fig:our-methods}
\end{figure*} 
\subsection{Spatial and graph structure}
\label{subsec:setup-notations}

\textbf{Traffic flow notations.}
Our traffic flow models assume that data is recorded at individual stations, and reported at time intervals of equal length $d$ by each station.
We denote by $F(s; i)$ the vehicle flow recorded by a station $s$, during the $i^{th}$ time interval.
Furthermore, because the stations are physically positioned along motorway segments (see \cref{Fig2_1_Methodology}a), they are of three types:
stations placed at the motorway entries and exists (denoted as $e$ and $x$ respectively), and stations placed along the main motorway (denoted here as $a$).
Stations also have a geographical relation (e.g. $s_1 \rightarrow s_2$) when they are consecutive along the motorway.
For example in $s_1 \rightarrow s_2$, a car that is recorded by $s_1$ could be recorded next by $s_2$.
Note that $s_1$, $s_2$ could be of type $a$, $e$ or $x$, with at least one of type $a$.

\textbf{Graph structure.}
We encode the geographical relations between stations into a graph structure of the stations $G(V, E)$ (depicted in \cref{Fig2_1_Methodology}b).
$V$ is the set of vertexes (or nodes) -- here the stations.
$E$ is the set of edges (or arcs) -- the geographical relation $s_1 \rightarrow s_2$, with $s_1, s_2 \in G$.
The edges are uni-directional as the physical motorway has separate ways for upstream and downstream traffic
The edges also have attributes, in our case we record the geographical distance between two stations, and the typical time it takes a vehicle to cover this distance at a constant speed of $90$km/h.

\textbf{Spatial station configurations.}
Within the graph $G$, we identify three station spatial structures of interest, showcased in \cref{Fig2_1_Methodology}c using real traffic flows from the dataset described in \cref{3_Case_study}.
The first structure is \textbf{Passing traffic}, consisting of two consecutive stations of type $a$, with no entries or exits between them.
\cref{Fig2_1_Methodology}$c_1$ shows the relation $19A \rightarrow 18A$, which follows the equation $F(19A; i) = F(18A; i) \pm \epsilon$, where $\epsilon$ accounts for the time cars take to move between stations, and for the inherent detector equipment error.
The second structure is \textbf{Exiting traffic}, containing one exit station.
\cref{Fig2_1_Methodology}$c_2$ shows that the traffic of station $4A$ splits into the traffic recorded by $3A$ and the exit $3X$, i.e. $F(4A; i)=F(3A; i) + F(3X; i) \pm \epsilon$.
The red line shows the sum of the two downstream stations $3A$ and $3X$, which matches closely the flow of station $4A$ (shown in green).
Finally, the third structure is \textbf{Merging traffic} and contains entry stations, shown in \cref{Fig2_1_Methodology}$c_3$.
Here the traffic from $12A$ and the entry $12E$ (their summed traffic shown in red) merge into the traffic recorded by $11A$ (shown in blue), i.e. $F(11A; i) = F(12A ; i) + F(12E; i)\pm \epsilon$.
Again we observe a tight overlap in the observed data corresponding to the two sides of the above equation, which motivates our graph flow prediction models detailed next.

\subsection{Backtracking prediction method (BKTR-P)} 
\label{subsec:historical-data}

Our backtracking prediction approach builds on the three spatial configurations of stations described in \cref{subsec:setup-notations} and on the following observation: 
once a vehicle is recorded by a given station, it has no other option but to travel downstream along the graph and be recorded at the following motorway stations (of type $a$), or at an exit station (type $e$).
Therefore, to predict the flow for a station in the future, it suffices to analyse the past recorded flow at upstream stations, adjusted for entries and exits.

\textbf{One step prediction.}
Formally, we aim to predict the flow for a station $v$ at $p$ time intervals in the future from the current time $t$, i.e. $\widehat{F}(v, t+p)$,
using the flow from $r$ intervals in the past $F(\cdot, t-r)$ -- here the dot is a placeholder for any station.
We identify the station $u$ of type $a$ which lies approximately $r+p$ time intervals upstream of station $v$ along the motorway.
This can be easily achieved by building a distance matrix between each two stations, and performing a lookup -- a vehicle will cover an expected distance of $(r+p) \times d \times \overline{speed}$ during $r+p$ time intervals.
Next, we traverse the graph structure between $u$ and v
and, in doing so, we identify the set of entries $\mathcal{E}$ and the set of exits $\mathcal{X}$ that a vehicle could take between stations $u$ and $v$.
Finally, to predict the flow at station $v$ one step into the future ($p=1$) we use the flow at $u$ one step in the past ($r=1$), to which we add the recorded flow of entries and we subtract the flow of exits:
\begin{equation} \label{eq:backtrack-1}
	\widehat{F}(v; t+1) = F (u; t-1) + \sum_{e \in \mathcal{E}} F (e; t-1) -\sum_{x \in \mathcal{X}}  F (x ; t-1)
\end{equation}
\noindent where $F(u ; t-1)$ denotes the flow at station $u$ during the last time interval before the present.

\textbf{Multi step prediction.}
When predicting more than one time step into the future, \cref{eq:backtrack-1} does not hold. 
This is because $\widehat{F}(v; t+p)$ is a function of $F(\cdot; t+p-1)$, which is also in the future for $p > 1$, and therefore unobserved.
One could build $\overline{F} (\cdot, t+p-1)$ (the historical average for a given station at a given time interval) and assume that the future traffic flow across the motorway follows the historical patterns of the given time of day.
However, doing so ignores changes, disruptions and other stochastic events that might be occurring at the time of prediction.
Our solution is to go further in the past (i.e. $r > 1$) and upstream on the motorway to find actual recorded flow values.
We call this approach \emph{backtracking}.

\cref{subfig:backtrack-schema} illustrates the backtracking approach over multiple time periods for $r=2$ and $p=2$.
The graph structure of the network is shown on the right hand side of the figure.
The x-axis represents time, with being $t$ the current time.
For a random vehicle to be recorded by station $33A$ at time $t+2$ (i.e. $p=2$, shown in yellow), it will need to originate from $43A$ three time intervals earlier (at $t-r, r = 2$).
The purple rectangles and the arrow between them represent the positions of the vehicle as it passes through the different stations while travelling between the stations $43A$ and $33A$, at a constant speed. 
Visibly, the vehicle is recorded by the stations $43A$, $42A$, $41A$ and $40A$ during the first time interval ($t-2$, $r = 2$), and by $39A$, $38A$ and $37A$ during the second time interval.
At the present time $t$, the car is located in between $37A$ and $36A$.
Two time intervals into the future, it is projected to reach $33A$ (the station for which we predict the flow).
Given the inputs $r = 2$, $p = 2$, the backtracking algorithm will estimate the flow at the station $33A$ at time interval $t+p$ using the values from $43A$ with a lag of $r+p$ time intervals, to which it adds the entries and subtracts the exists at the corresponding time intervals (here $+ F(41E; t-2)$, $+F(38E; t-1)$ and $-F(39X; t-1)$).
In \cref{subfig:backtrack-schema}, it adds and subtracts the areas in light blue, and ignores the areas in red.
For any entries and exits occurring in the future, we use the historical mean, e.g. $\overline{F} (\cdot, t+1)$. 

\begin{algorithm}[tbp]
	\caption{
		\textbf{BTRK-P:} predict traffic flow using backtracking, for a single station on one day.
	}
\begin{algorithmic} 
	\REQUIRE $G(V, E)$ -- graph of stations along motorway;\\
			 $V$ -- set of stations; $E$ -- set of geographical relations between stations\\
			 $p$ -- number of steps in the future to predict flow;\\
			 $r$ -- number of time intervals in the past to use;\\
			 $v$ -- station for which to predict the flow;\\
			 $F(\cdot; \{t-r, t-r+1, .. , t-1\})$ -- observed flow for all stations from $t-p$ until present;\\
			 $\overline{F}(\cdot; \{t+1, .. , t+p-1\})$ -- historical means for all stations at future time points.
	\ENSURE $\widehat{F}(v ; t+p)$ -- traffic flow for station $v$ at $p$ intervals in the future.
	\STATE \textit{\textcolor{gray}{// find optimal $u$ for target $v$;}}
	\STATE $optim\_dist \leftarrow (r+p) \times d \times speed$ ;
	\STATE Find $u \in V$ of type $a$ so that 
	$len(u \rightarrow s_k) + \sum_{j=k}^{l-1} len(s_j \rightarrow s_{j+1}) + len(s_l \rightarrow v) \approx optim\_dist,$ with $s_k, .., s_l \in V$ ;
	\STATE \textit{\textcolor{gray}{// traverse from $u$ to $v$ to identify entries and exits;}}
	\STATE $\mathcal{E} \leftarrow \emptyset$; $\mathcal{X} \leftarrow \emptyset$;
	\STATE $\mathcal{B} \leftarrow \{u\}$; \textit{\textcolor{gray}{// Breadth First Search node set}}
	\STATE $dist\_from\_u \leftarrow 0$;
	\WHILE{$\mathcal{B} \neq \emptyset$}
		\STATE pop $head$ from $\mathcal{B}$;  
		\IF{$head$ of type $a$}
			\STATE \textit{\textcolor{gray}{// update the distance from $u$}}
			\STATE $dist\_from\_u \leftarrow dist\_from\_u + len(head, p(head))$
			\STATE \textit{\textcolor{gray}{// add stations linked to $head$ to the queue}}
			\STATE \textit{\textcolor{gray}{// first those of type $a$, then entries and exits}}
			\STATE $\mathcal{B} \leftarrow \mathcal{B} \cup \{s\}$, $\forall s \in V$ with $head \rightarrow s, type(s) = a$
			\STATE $\mathcal{B} \leftarrow \mathcal{B} \cup \{s\}$, $\forall s \in V$ with $head \rightarrow s, type(s) \in \{e, x\}$
			\STATE $p(s) \leftarrow head, \forall s \in V$ with $head \rightarrow s$
		\ENDIF
		\IF{$head$ of type $e$}
			\STATE \textit{\textcolor{gray}{// add $head$ to entries group}}
			\STATE $\mathcal{E} \leftarrow \mathcal{E} \cup \{head\}$;
		\ENDIF
		\IF{$head$ of type $x$}
			\STATE \textit{\textcolor{gray}{// add $head$ to exits group}}
			\STATE $\mathcal{X} \leftarrow \mathcal{X} \cup \{head\}$;
		\ENDIF
		\STATE compute $align(head)$ on time intervals based on $dist\_from\_u$;
	\ENDWHILE	
	\STATE \textit{\textcolor{gray}{// compute predicted flow for $v$;}}
	\STATE $\widehat{F}(v ; t+p) \leftarrow F(u; t-r)$
\STATE \textit{\textcolor{gray}{// add traffic of entries}}
	\FOR{each $e \in \mathcal{E}$}
		\IF{$align(e) < p$}
			\STATE $\widehat{F}(v ; t+p) \leftarrow \widehat{F}(v ; t+p) + F(e; align(e))$
		\ELSE
			\STATE $\widehat{F}(v ; t+p) \leftarrow \widehat{F}(v ; t+p) + \overline{F}(e; align(e))$ 
		\ENDIF
	\ENDFOR
	\STATE \textit{\textcolor{gray}{// subtract traffic of exits}}
	\FOR{each $x \in \mathcal{X}$}
		\IF{$align(e) < p$}
			\STATE $\widehat{F}(v ; t+p) \leftarrow \widehat{F}(v ; t+p) - F(x; align(x))$
		\ELSE
			\STATE $\widehat{F}(v ; t+p) \leftarrow \widehat{F}(v ; t+p) - \overline{F}(x; align(x))$ 
		\ENDIF
	\ENDFOR
	
\RETURN $\widehat{F}(v ; t+p)$
	\end{algorithmic}
	\label{algo:backtrack}
\end{algorithm} 
The backtracking algorithm (summarised in \cref{algo:backtrack}) aligns entries and exits to exact time intervals using the distance from station $u$, alongside with the length between stations and the average travel speed. We make the observation that $p(head)$ stands for the predecessor of the ``head'' node extracted from B, which is the ``Breadth First Search node set'' (a computer science typical graph-traversal algorithm). It then chooses the time interval closest to the computed time interval. Obviously, the travel time between stations does not align with the time intervals. Therefore, we also tested another method to estimate the traffic at these stations based on observed traffic and historical means to see whether this would further improve efficiency of our proposed algorithm or not.

\subsection{Interpolation prediction using actual and historical data (INTR-P)}
\label{subsec:models}

\textbf{Interpolating flow across two stations.}
We further test a method to estimate the number of vehicles that pass through a station by interpolating the recorded values in two consecutive time intervals -- denoted as $F^*(s; i \rightarrow i+1)$.
\cref{subfig:interpolation-explain} shows how the interpolation is achieved.
We aim to compute the flow for a station overlapped over two time intervals, which covers the last $t_1$ seconds from the first time interval and the first $t_2$ seconds from the second interval.
Note that $t_1 + t_2 = d$ (with $d$ the length of time intervals), so we compute the interpolation over a length $d$ shifted from the reported values.
We assume that vehicles arrive uniformly distributed in front of the station detectors. Consequently, during the time $t_1 < d$ there are $\dfrac{t_1 F(s; i)}{d}$ vehicles passing by (where $F(s; i)$ is the recorded flow at the $i^{th}$ interval).
Similarly, during the first $t_2$ period of the second interval $i+1$ pass $\dfrac{t_2 F(s; i+1)}{d}$ vehicles.
Finally, we estimate the interpolated traffic flow at station $s$ as:
\begin{equation} \label{eq:interpolation}
	F^*(s; i \rightarrow i+1) = \dfrac{t_1 F(s; i) + t_2 F(s; i+1)}{d},\text{ with }t_1 + t_2 = d
\end{equation}

\cref{subfig:interpolation-schema} revisits the previous example for $r=2, p=2$ discussed in \cref{subfig:backtrack-schema,subsec:historical-data}.
The purple rectangle again represents a random vehicle and is its position when travelling from $43A$ to $33A$ at constant speed. 
Given the interpolation described above, the areas that we add or substract can overlap over two time intervals (see $41E$, $39X$ and $38E$).
Visibly, for the intervals not contributing to the computation (red areas of main motorway), the vehicle now stays at the same position in the interpolated flow segment. 
In the original backtracking example (\cref{subfig:backtrack-schema}), the vehicle changed from being at the back to being closer to the front of the flow until eventually returning to the initial position on the predicted flow. 

\textbf{The interpolation algorithm (INTR-P).}
The interpolation method is a variation of the backtracking algorithm.
Just like \cref{algo:backtrack}, it searches for the optimal station $u$ which is located $(r+p) \times d \times speed$ distance away from $v$.
However, when it adds and subtracts the flow of entries and exits, it uses the interpolated flow $F^*(s; i \rightarrow i+1)$ instead of the actual flow.
There are two more changes for INTR-P over BKTR-P.
First, the times $t_1$ and $t_2$ are computed based on the physical distance between the station $s$, $type(s) \in \{e, x\}$ and station $u$.
Second, when the interpolated interval spans across an observed interval and a future interval, we use the historical mean for a particular day of the week in the second term of the right hand side of \cref{eq:interpolation}.
The overall algorithm for INTR-P is shown in the online supplement~\citep{appendix} due to space constraints.

\section{Case study}\label{3_Case_study}

\subsection{Description}\label{S31_Description}

The current work employs the same motorway traffic flow dataset as our prior work \citep{Mihaita2019}.
The dataset was collected over the entire year of 2017, by recording the traffic flow at each of the 208 bi-directional ``flow counting stations'' along the M7 Motorway in Sydney, Australia (see \cref{Fig2_1_Methodology}). 
The dataset contains 36.34 million data points, where one data point is the flow recorded by one station during time intervals of length $d = 3min$.
We denote as $d_i$ the $i^{th}$ interval of a day, $i = 1..480$.

\subsection{Data profiling}\label{S32_Data_profiling}

\begin{figure}[tbp]
	\centering
\newcommand\mywidth{0.45} 

	\subfloat[]{
		\includegraphics[width=\mywidth\textwidth]{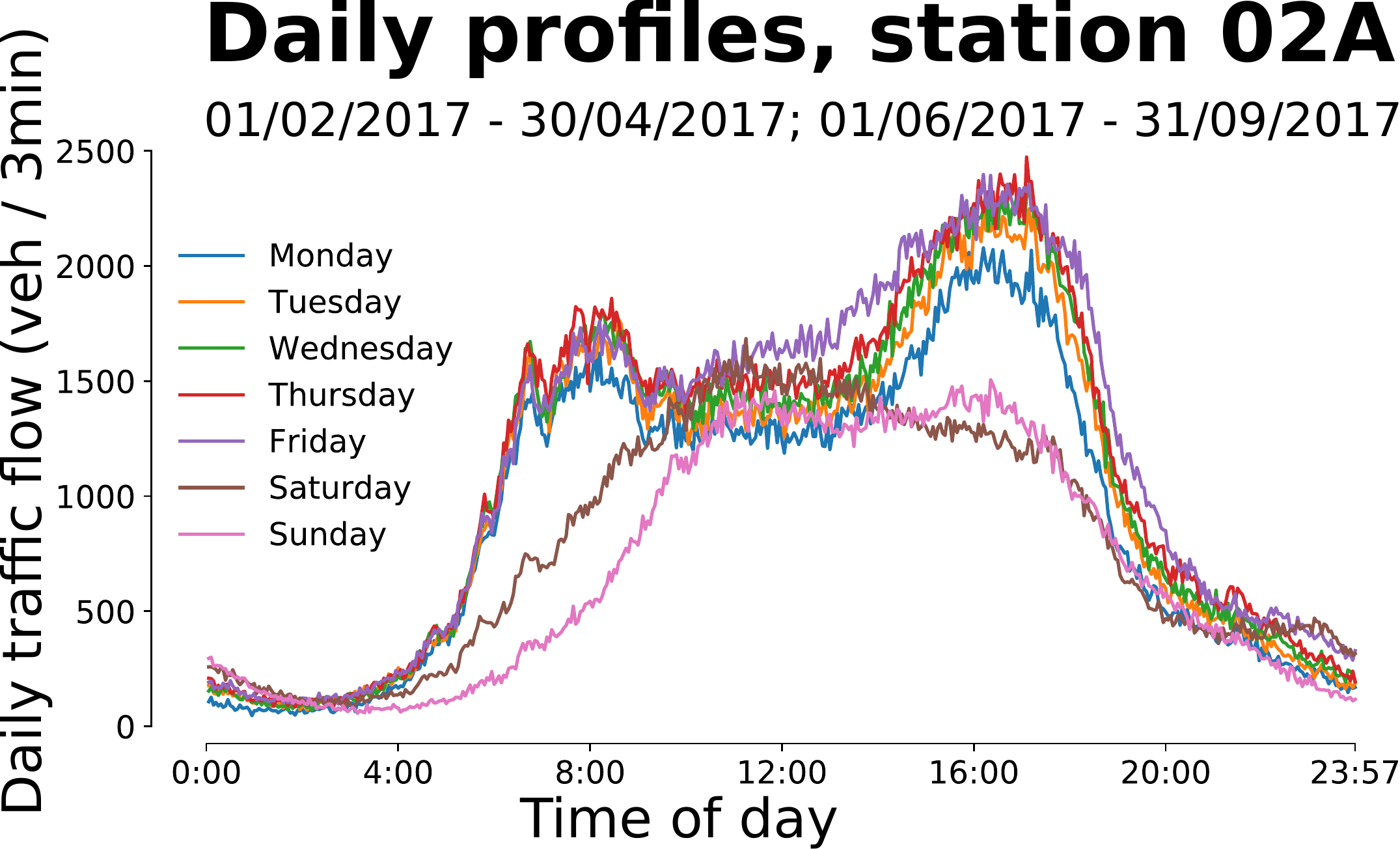}\label{subfig:DPP_by_day}}\\\subfloat[]{
	    \includegraphics[width=\mywidth\textwidth]{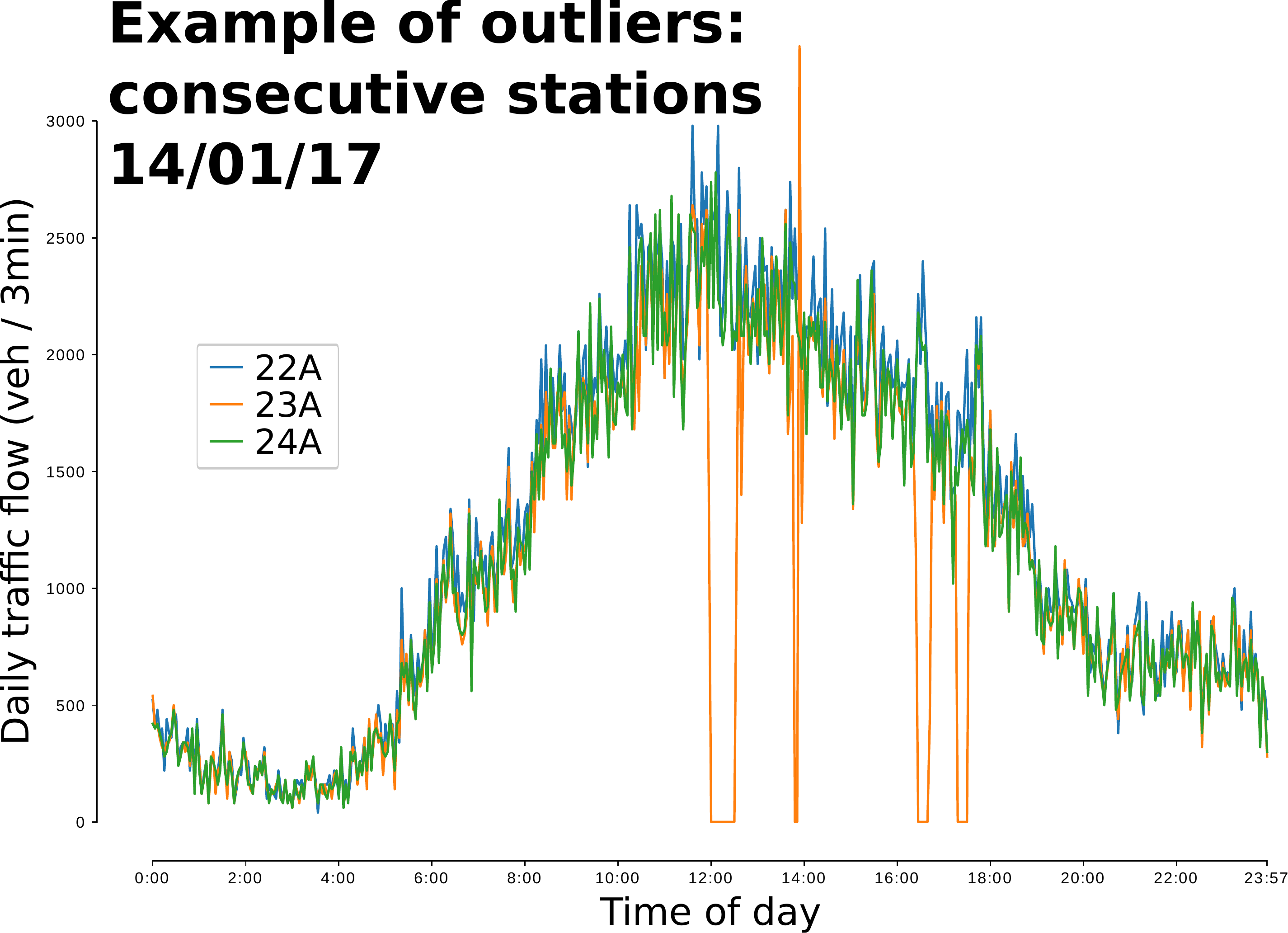}\label{subfig:Example_outliers}}
	\caption{a) Daily Profile - Historical data by day of the week, b) Example of traffic outliers and anomaly detection across consecutive traffic flow stations without any entries/exits in between.} 
		
	\label{fig:DPP_example_of_outliers}
\end{figure}

\textbf{The daily profile.}
We first observe that the traffic flow at any given station presents a strong daily and weekly seasonality, mainly driven by the users daily work commute patterns. We define the \emph{daily profile} as the typical daily traffic flow series recorded at a given station and we compute it as the average flow for each $d_i$, for a given day of the week, over a training period of 6 months from the study year.
\cref{subfig:DPP_by_day} shows the Daily profile extracted by each day of the week, which indicates two peaks corresponding to two rush hours for Monday-to-Friday days (one in the morning (8-10AM) corresponding to the daily commute towards work, and a second one in the afternoon (5-8PM) for returning from work), and a single peak for weekend between 11AM-4PM.

\textbf{Outlier analysis and correction.}
One immediate usage for the daily profile is to identify the non-standard days such as public holidays or special-event days.
Moreover, the daily profile can help identify traffic anomalies and other traffic volume outliers. 
\cref{subfig:Example_outliers} presents the example of stations $22A$, $23A$ and $24A$, all three consecutive stations of type $a$ arranged in the passing traffic spatial configuration (see \cref{Fig2_1_Methodology}$c_1$).
As there are no entries and no exits in between, the traffic flow for the three example stations is expected to follow similar patterns. 
However, the middle station $23A$ seems different from its neighbours $22A$ and $24A$, and has several drop-outs (between 12PM-17PM) together with a high peak around 2PM -- which we interpret as anomalies in the traffic flow.
We remove such outliers by leveraging the method proposed by \citet{Mihaita2020}.
In a nutshell, we use the moving average of the deviation from the daily profile, alongside with the recent observed traffic volumes to identify and impute outliers.
All results presented in this paper are obtained on the cleaned data set, after removing outliers and anomalies.

\section{Experimental set-up}\label{S4_Experiments}

In this section we describe the set-up and the implementation of the two proposed graph-based prediction methods, as well as the comparison with other deep learning method previously developed for the same motorway network. 

\paragraph{\textbf{Prediction setup}}
\label{D4_experiment_setup}
To be able to compare to previously reported results using deep neural approaches, we train the BKTR-P and INTR-P models
on 7 months of the entire dataset.
\citet{Mihaita2019} selected this period to avoid missing and erroneous data.
More specifically, we use the traffic flow from February-April and June-August 2017 respectively, for training the models (6 months in total). 
The flow from October 2017 is used for testing.
This was kept for consistency across all the DL models that we have retrained. 	

\paragraph{\textbf{Past and future prediction horizon selection}}
 
$r$ is an important hyper-parameter of our graph-models (the length of the learning past horizon) which is tuned on the validation set in the range $R \in\{1,..5\}$, where 5 points in the past correspond to a 15 min past horizon. Similarly, $p$ represents the future prediction points to estimate the traffic flow and we also fluctuate it the range $p \in\{1,..5\}$. The main constraint when applying the above graph-based models relies mainly on the number of stations that we can use for the prediction.
We are bounded by the lengths of the motorway station, since we need to go farther upstream on the motorway to produce predictions further in time.
 
For example when considering the case of {$r=1$, $p=1$} (predicting 3 min in the future by using the traffic flow of all closest counting stations from last 3-min time interval), the BKTR-P and INTR-P prediction models can be applied on almost all stations, from $072A$ until $03A$.
However, when we increase either the past or future horizon that we want to use for the prediction (say $r=5, p=5$ standing for predicted flow 15-min in the future using latest 15-min in the past), we can only apply the graph modelling predictions for stations from $26A$ to $03A$,
This is mainly due to limited dimension of the highway and time and space constraints. 
 
For consistency, we keep the same range of variation when comparing with other deep learning/ baseline models ($r,p\in \{1,..5\}$) and the results will be discussed in \cref{S53_DL_comparison}. The main objective of this experiment is to learn the efficiency of the proposed models in time and space and answer some challenging questions such as: 
a) how much should we learn from the past to achieve best prediction results when using graph-oriented models versus regular DL models? 
b) how long in the future can we predict accurately using graph-modelling? 
c) what is the best past and future combination that can achieve best prediction results?

\paragraph{\textbf{Other baseline models}}

we compare the performance metrics of the proposed graph models to other models such as:
 \textbf{Daily Profile Prediction (DPP)} and various state-of-art DL models. 
 DPP is a base model in which we use the Daily Profile (see \cref{S2_Method}) computed for each station and each day of the week as a predictor; this represents basically the prediction of the average traffic flow counts by each day of the week which we call “daily profiles”. We use this as a reference to any prediction comparison and we believe that any prediction doing worse than predicting the “historical average traffic profiles” of that particular day is basically a bad predictor. For example, for each counting station, we have 7 flow curves and each curve is consisting of 480 flow values (the time interval between counts is 3min). This model therefore predicts the average traffic flow per station. We also used our previously developed deep learning models on the same network structure and available data set, such as: back-propagation neuronal networks (\textbf{BPNN}), convolutional neuronal networks (\textbf{CNN}), long-short term memory models (\textbf{LSTM}) and the hybrid \textbf{CNN-LSTM}. The deployment, implementation, hyper-parameter tunning of these DL models has been published in \citep{Mihaita2019}. All our DL models are implemented in PyTorch~\cite{paszke2017automatic}, using the Adam optimiser which provided a better performance than SGD or AdaGrad.

 \paragraph{\textbf{Model performance}}
 We evaluate prediction performances using three widely used measures such as:
 
 i) the Root Mean Square Error:
 \begin{equation*}
     RMSE=\sqrt{\dfrac{1}{N} \sum_{i=1}^N \left(\widehat{F}(v;t)-F(v;t)\right)^2}
 \end{equation*}
 
  ii) the R-squared value:
  \begin{equation*}
  R^2 = 1 - \dfrac{\sum_{i=1}^N \left( \widehat{F}(v;t) - {F}(v;t)\right)^2}{\sum_{i=1}^N \left(  {F}(v;t) -  \overline{F}(v;t) \right)^2  } 
   \end{equation*}

  iii) Symmetric Mean Absolute Percentage Error:
  \begin{equation*}
  SMAPE = \frac{100\%}{n}\sum_{i=1}^N \frac{| \widehat{F}(v;t)-{F}(v;t) |}{|\widehat{F}(v;t)| + | {F}(v;t)|}.
  \end{equation*}
 
\section{Results}\label{S5_results}
	\subsection{Anomaly and outlier treatment}\label{S51_anomaly_detection}
	
	As mentioned in \cref{S32_Data_profiling}, the data profiling procedure revealed several missing data points throughout the stations, outliers (very large or very small vehicle counts compared to the daily flow trend) or anomalies (very low flow counts due to public events or traffic disruptions). These phenomena led to the outlier and anomaly detection algorithm that we applied for cleaning the data sets before using it for graph-prediction modelling which can be sensitive to such abnormal behaviour. As an example, if the prediction of a station relies purely on its closest preceding stations which are affected by abnormal events, then the predicted traffic flow would be also affected.
\begin{figure}[t]
    \centering
	\includegraphics[width=0.48\textwidth]{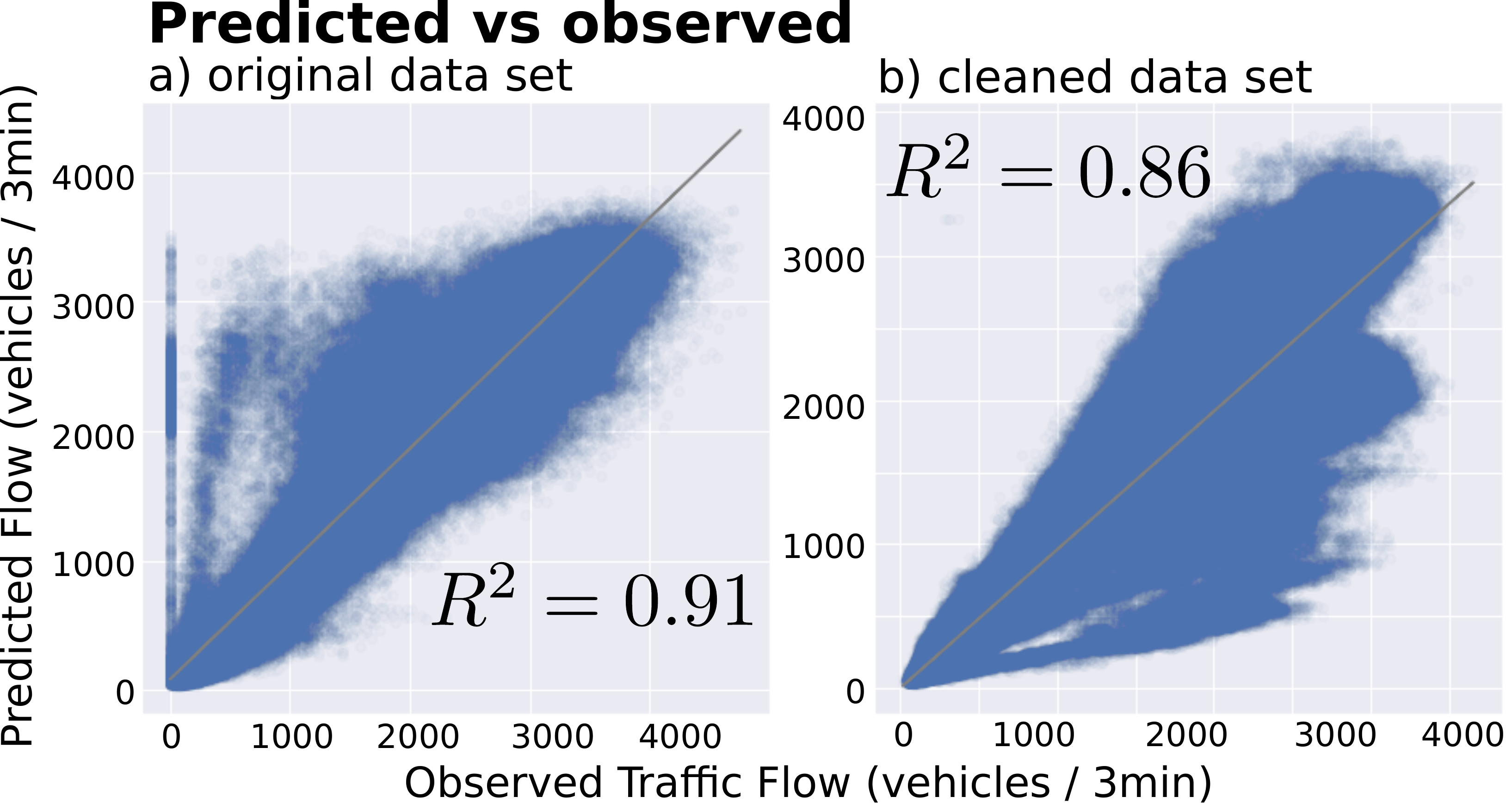}
	\caption{Daily Profile Predictor before and after data cleaning.}
	\label{Fig_5_1_DPP_before_after}
\end{figure}

\cref{Fig_5_1_DPP_before_after} showcases the $R^2$ results for DPP, before and after appplying the cleaning, which we considered as an initial baseline model for future model comparison. Although when using original data sets, $R^2$ recorded a 0.91 value, there were many zero flow values (and small values less than $500 veh/T_i$) being over predicted (see \cref{Fig_5_1_DPP_before_after}a)), whereas after cleaning, although $R^2$ dropped to 0.86, it obtained better fitting and reduced significantly the chances of anomalies. Similar performance improvements have been observed when analysing RMSE and SMAPE results, which are further provided in our online supplement~\cite{appendix} (due to lack of space).  
	  
\subsection{BKTR-P and INTR-P results}\label{S52_anomaly_detection}
\subsubsection{Short-term prediction results (r=1,p=1)}\label{S521_r1p1}
	
\begin{figure}[tbp]
	\centering
\newcommand\mywidth{0.45} 

	\subfloat[]{
		\includegraphics[width=.35\textwidth]{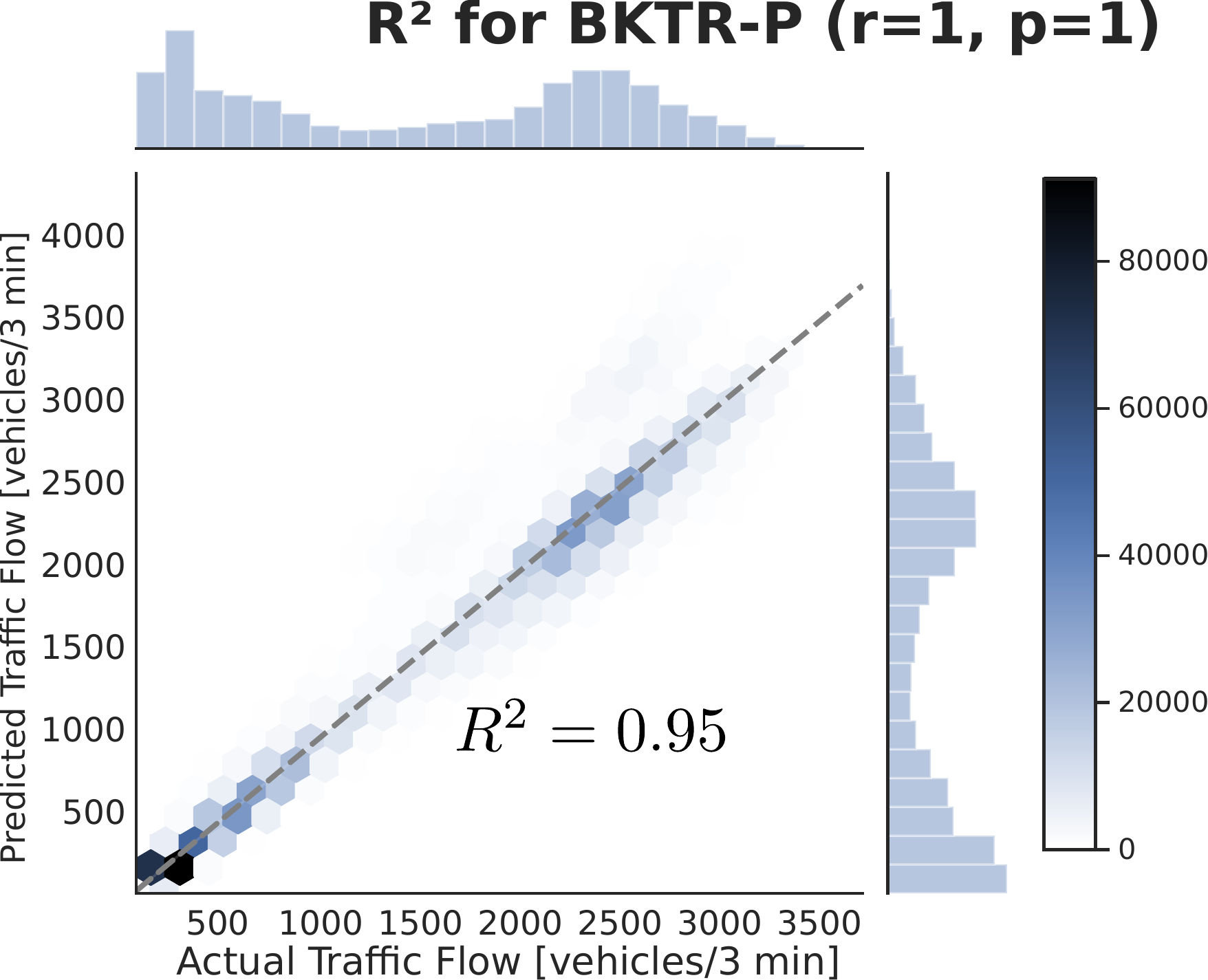}\label{subfig:BKTR_R2_r1p1}}
	\\
	\subfloat[]{
	    \includegraphics[width=\mywidth\textwidth]{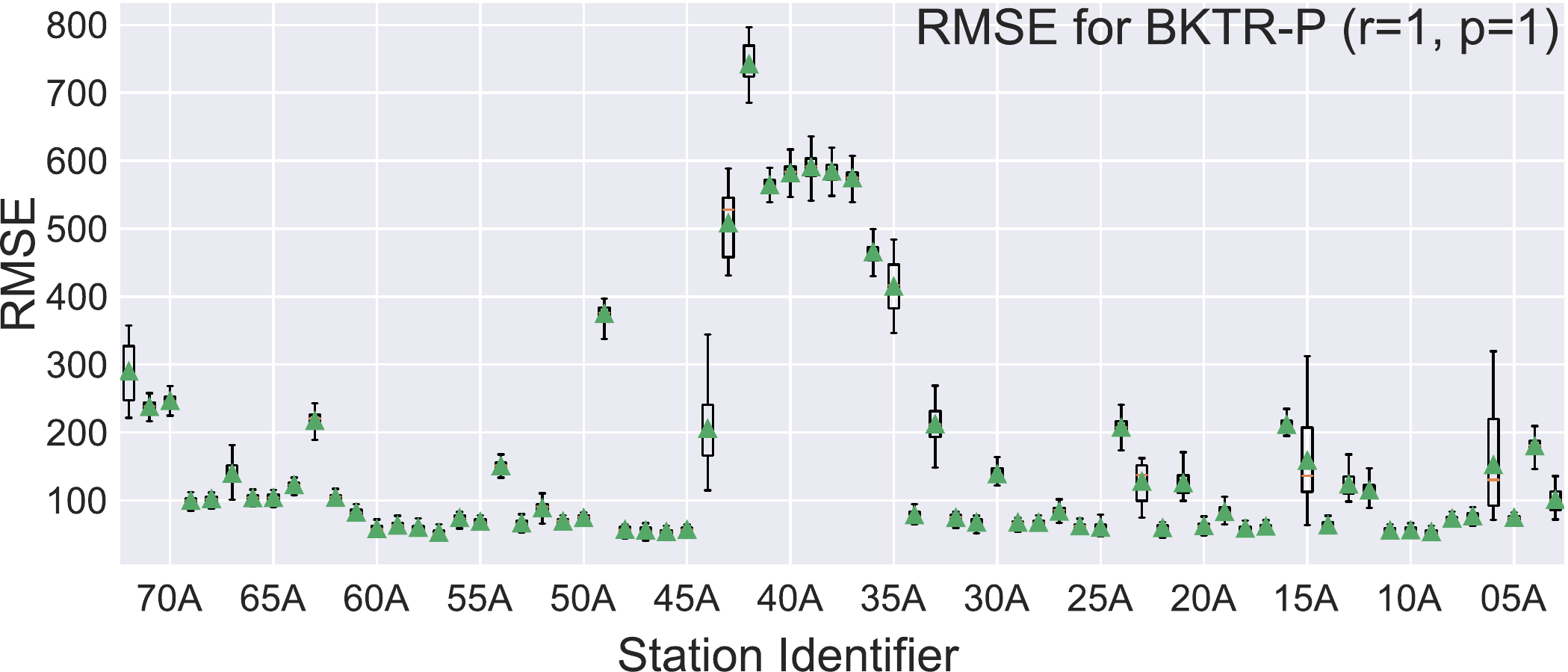}\label{subfig:BKTR_RMSE_r1p1}}
	\\
	\subfloat[]{
		\includegraphics[width=\mywidth\textwidth]{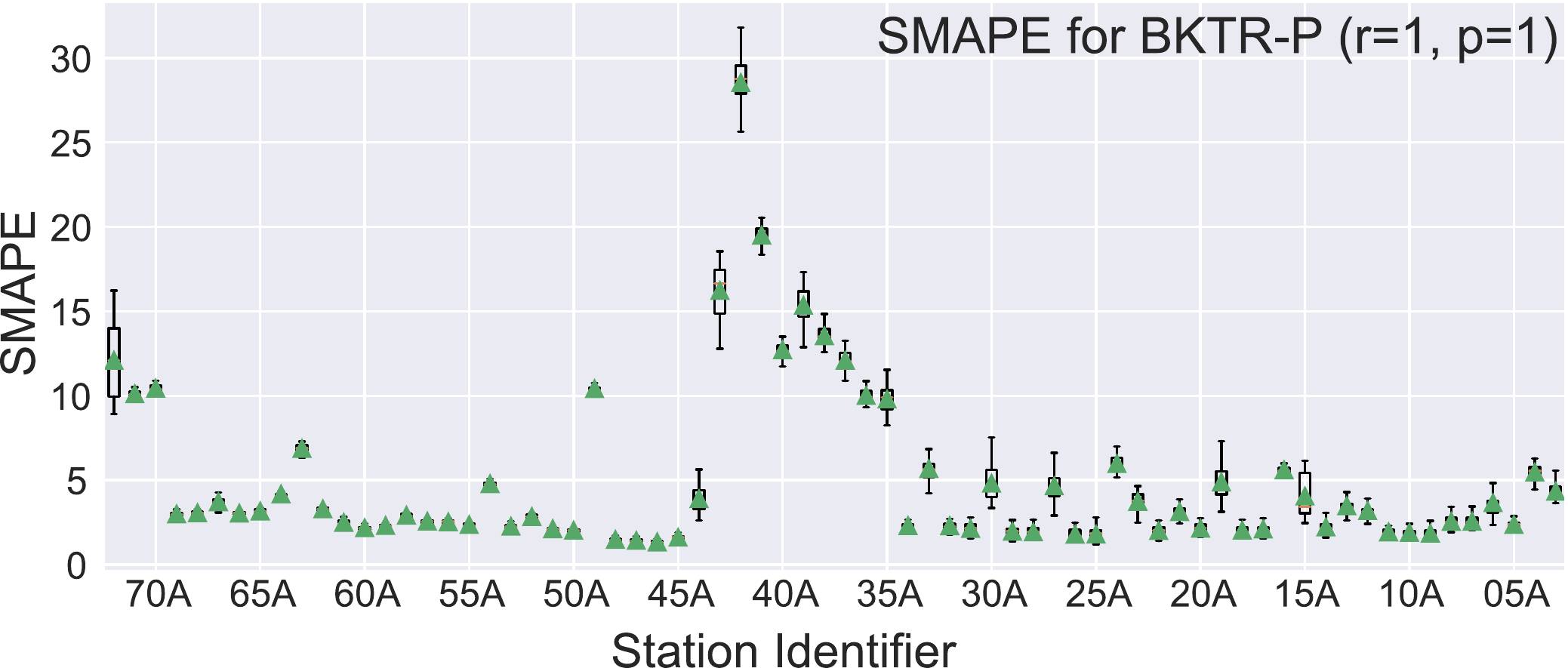}\label{subfig:BKTR_SMAPE_r1p1}}
	\caption{
		\textbf{Traffic flow prediction using BKTR-P for all direction $A$ stations.}
		\textbf{(a)} The $R^2$ coefficient of determination between the actual traffic flow (x-axis) and the predicted flow (y-axis).
		\textbf{(b)(c)} RMSE (b) and SMAPE (b) prediction errors for each station, as boxplots.
		The green triangles show average values.
	} 
		
	\label{fig:BKTR_INTRP_metrics}
\end{figure}

We now analyse the performance of the two proposed methods: BKTR-P and INTR-P, across a wide range of stations ($71A-03A$) for a short time prediction horizon $p=1$ (3min in the future) using the traffic flow information from the last 3min in the past ($r=1$). \cref{fig:BKTR_INTRP_metrics} showcases the performance results for the BKTR-P method such as: the $R^2$ value, this time expressed as density plots for a better visualisation, followed by the RMSE and SMAPE across all stations, expressed as box plots to showcase the variation of prediction results across all stations. More specifically, \cref{fig:BKTR_INTRP_metrics}a) reflects an $R^2$ of 0.95, with two main density regions based on the incoming flow on the motorway (one for flows less than 1000 veh/3min and another one between 2000-3000 veh/3min) which reflects a good adaptation of the BKTR-P to the daily vehicle flow trend. 
The RMSE values range around 80-90 (compared to 280-300 for DPP - see the supplement ~\cite[Fig. 8c]{appendix}) while SMAPE reaches as well very good accuracy of $3-4\%$ across majority of stations (compared to DPP SMAPE values which were double or triple on majority of stations as shown in supplement ~\cite[Fig. 8e]{appendix}). We also make the observation the predicted flows for stations between $45A$ and $35A$ using BKTR-P present higher errors than the rest of the motorway; this is mainly due to the interconnection of M7 motorway with the M4 motorway (often affected by congestion and accidents at its merging/diverging with other motorways). 

Similarly, INTR-P provides as well good prediction results in terms of $R=0.94$ (see supplement ~\cite[Fig. 9a]{appendix}), RMSE ranging from 100-130 (see supplement~\cite[Fig. 9b]{appendix}) and SMAPE maintaining values between $3-5\%$ (see supplement~\cite[Fig. 9c]{appendix}). There is however, a slight better performance of BKTR-P compared to INTR-P which makes it the preferred graph-predictor for our case so far.

			\subsubsection{Varying past and future prediction horizons}\label{S521_r5p5}

\begin{figure}[t]
	\centering
\newcommand\mywidth{0.22} 

	\subfloat[]{
		\includegraphics[width=\mywidth\textwidth]{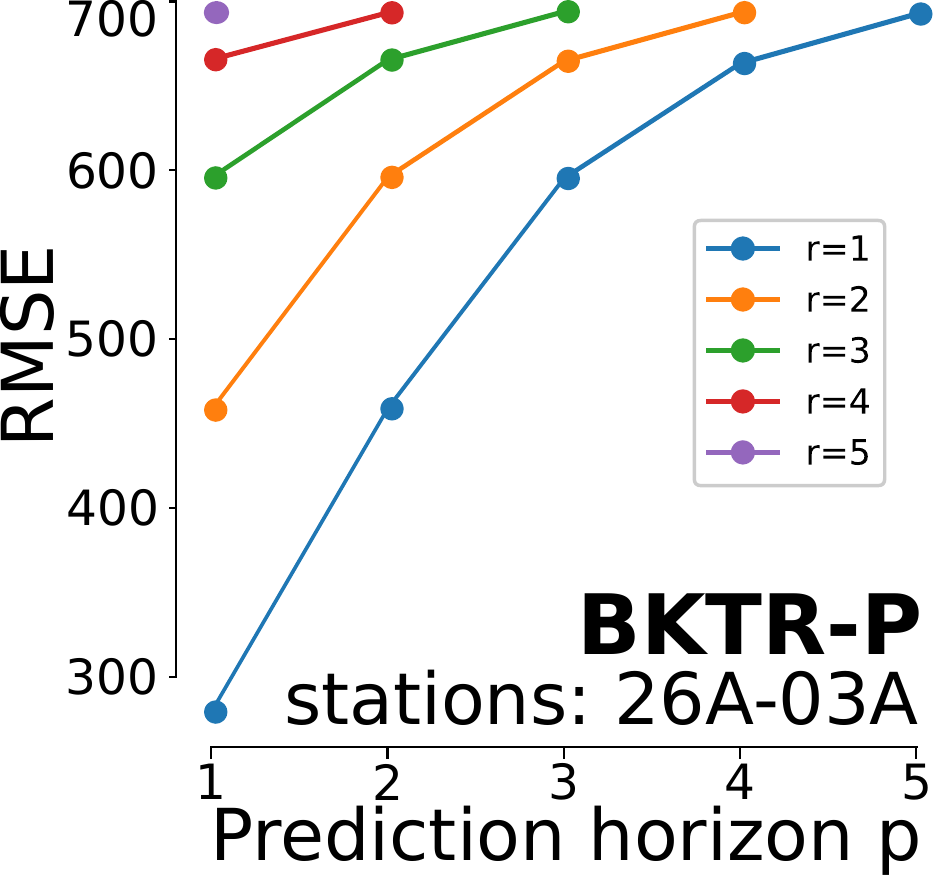}\label{subfig:BKTR_RMSE_rp15}}
	\subfloat[]{
		\includegraphics[width=\mywidth\textwidth]{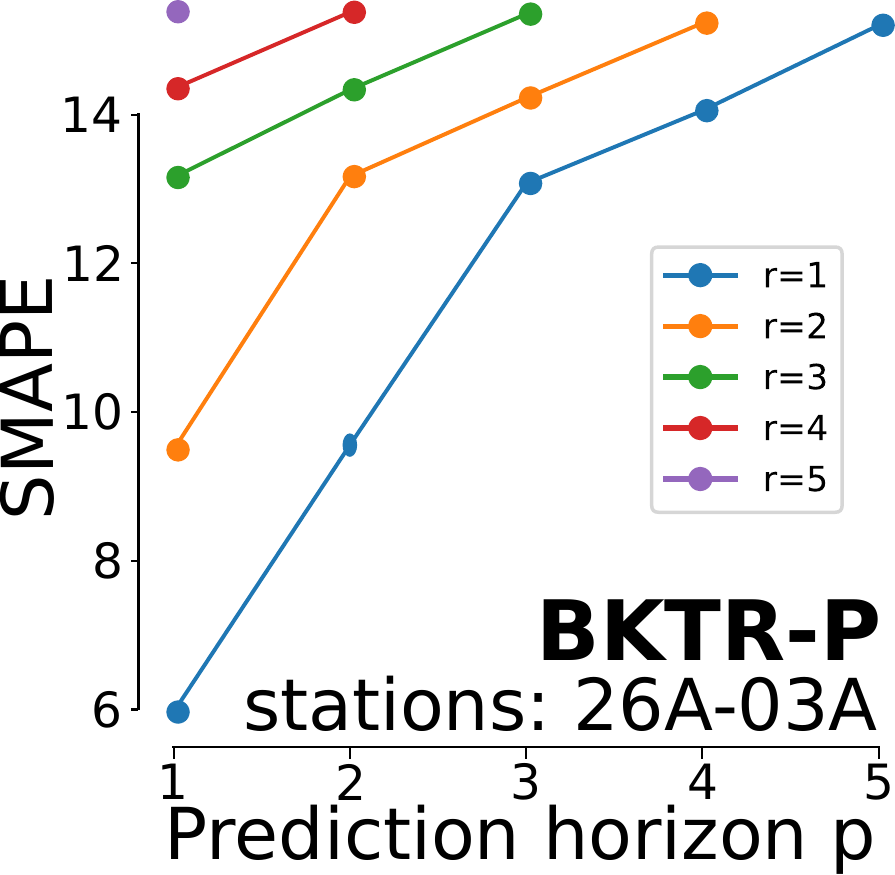}\label{subfig:BKTR_SMAPE_rp15}}
	\caption{BKTR-P prediction results for various past and future time horizons $r$, $p$: a) RMSE b) SMAPE.} 
		
	\label{fig:BKTR_P_Past_future_horizons}
\end{figure}

Moving further, as mentioned in \cref{S4_Experiments}b), a main question we want to answer is how much can we use from the past to achieve good prediction results when using graph-modelling prediction? Therefore, we have verified the performance of both models BKTR-P and INTR-P against variations of $p\in\{1,..5\}$ and $p\in\{1,..5\}$ and provided the results in \cref{fig:BKTR_P_Past_future_horizons} and ~\cite[Fig. 10a,b]{appendix}) respectively. For each variation of past and future horizon, both methods can be applied on a limited number of stations, and in this case, for $r=5, p=5$, the maximum number of stations that can predicted will be limited to $26A-03A$. 
The experiments have been carried by varying both parameters consecutively and the results are presented incrementally. BKTR-P seems to slightly overtake the INTR-P in terms of accuracy, achieving best results for example for short-term predictions ($p\leq 3, r\leq 3)$; when the algorithms tries to predict longer periods in the future ($p=5$), the relevance of past traffic flow information used for the prediction seems to decrease, providing almost similar results in terms of both SMAPE and RMSE; similar behaviour is present for INTR-P. This is mostly related to the nature and construction of the graph-based approaches in which the proximity between nodes and the travel time between them are more critical to how far along the branches you can predict in both time and space: for example, traffic flow circulating on far away branches of the graph are less likely to impact immediately the traffic flow at a current node.

\subsection{Comparison with other Deep Learning models}\label{S53_DL_comparison}
\subsubsection{DL comparison (r=1,p=1)}

\begin{figure}[tbp]
	\centering
	\newcommand\myheight{0.195} \subfloat[]{
	    \includegraphics[width=0.4\textwidth]{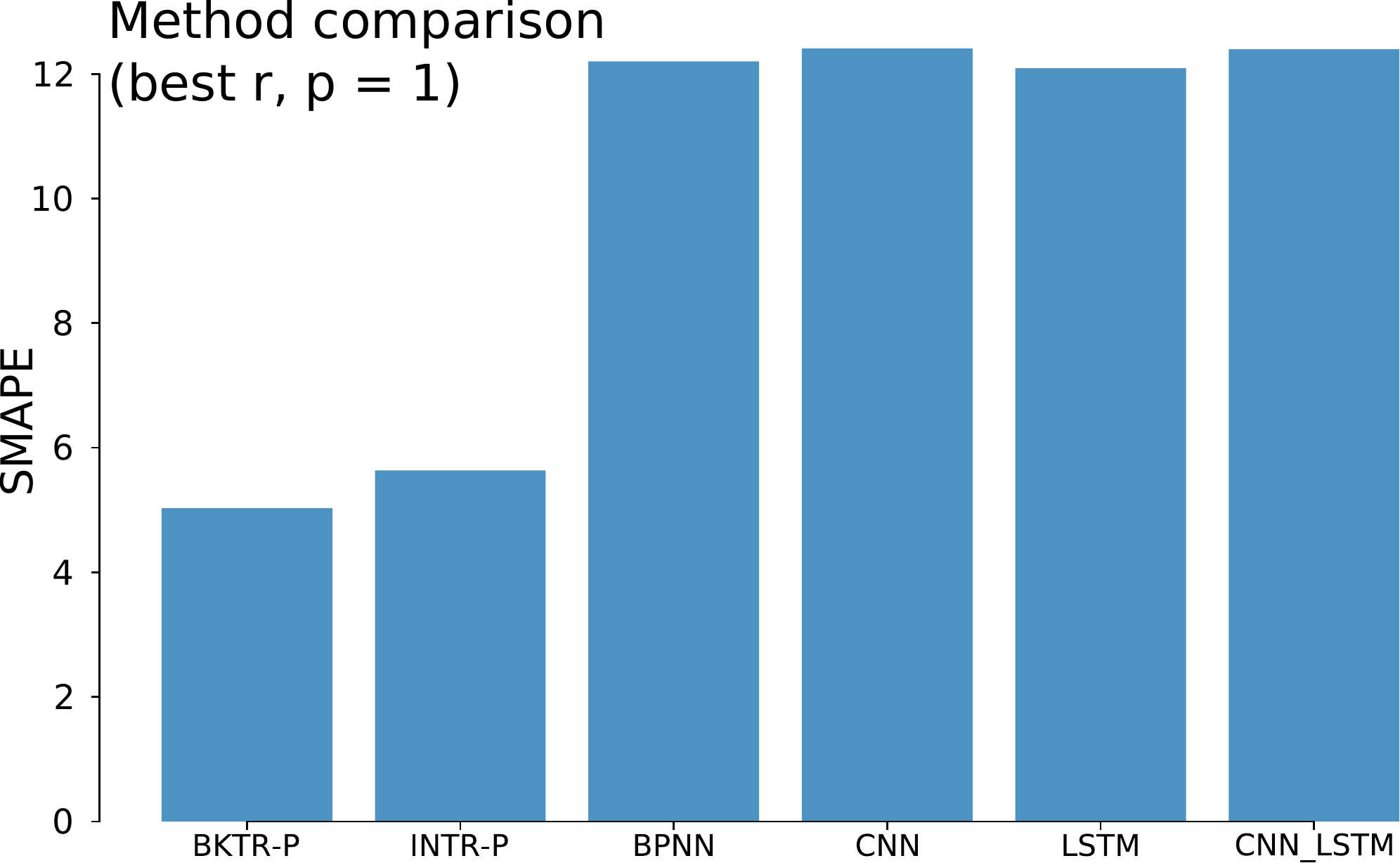}\label{subfig:SMAPE_All_methods_best_r_for_p1}}\\\subfloat[]{
		\includegraphics[width=0.4\textwidth]{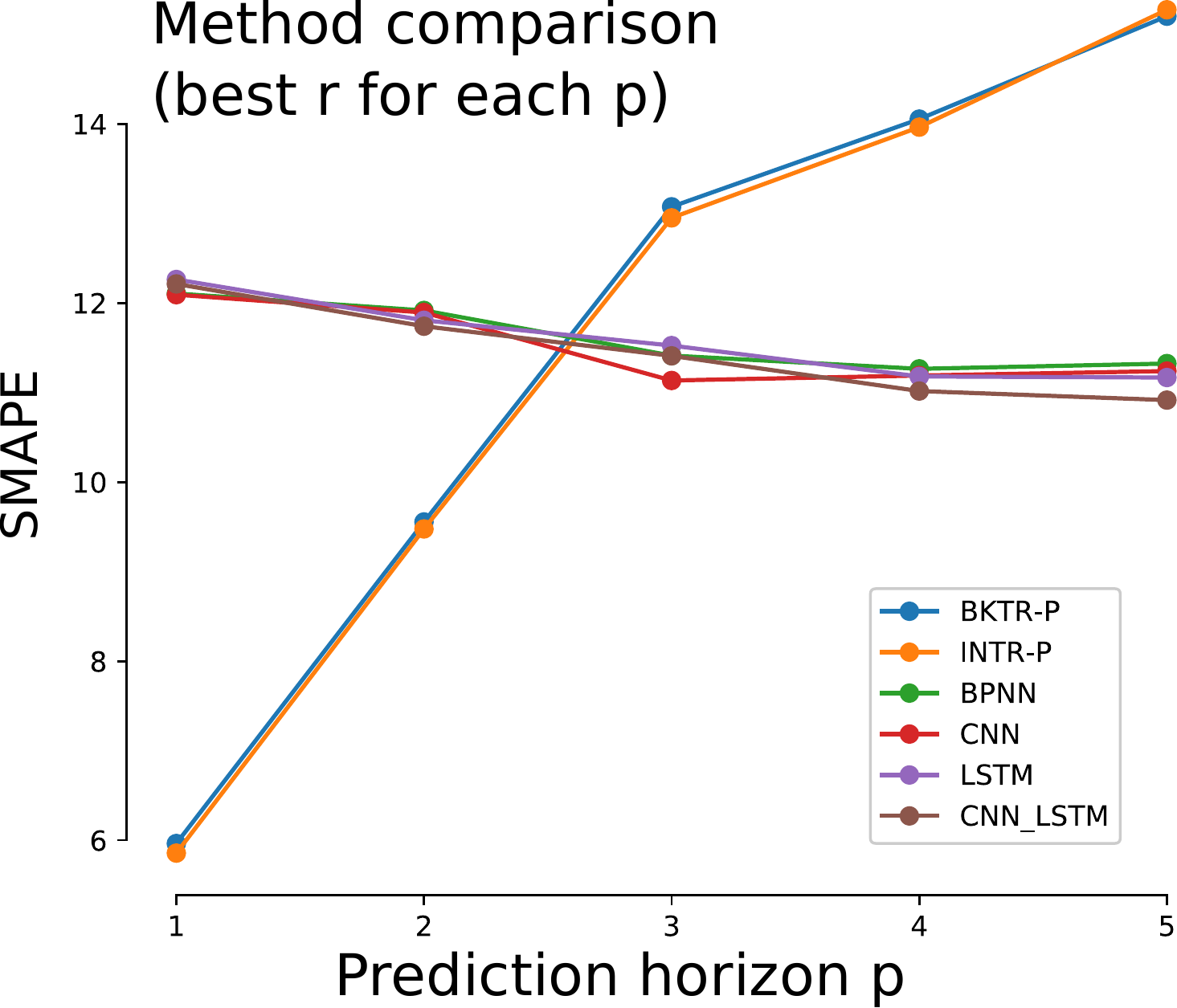}\label{subfig:SMAPE_All_methods_best_r_for_p5}}\caption{ 
		Traffic flow prediction for stations $71A$ -- $03A$ a) SMAPE using best r for p=1, b) SMAPE using best r for $p\in\{1,..5\}$. } 		
	\label{fig:SMAPE_Past_future_horizons_best_p5}
\end{figure}

In order to test the performance of the proposed models, we have further compared the results across other deep learning models developed for the flow prediction on M7 motorway. \cref{subfig:SMAPE_All_methods_best_r_for_p1} presents the comparison of   BKTR-P and INTR-P against BPNN, CNN, LSTM and the hybrid CNN-LSTM, for short term prediction ($r=1,p=1$) evaluated using SMAPE. The comparison is done by aggregating all prediction results across all stations $72A-03A$. BKTR-P seems to outperform all DL models, by almost $60\%$ as the previous best DL model LSTM achieved a SMAPE of $12\%$ compared to BKTR-P at $5\%$. Similar performance if obtained when analysing the RMSE (figures provided in supplement ~\cite[Fig. 11a)]{appendix}.
 			 
 			 \subsubsection{DL comparison (r=1..5,p=1..5)}
 			 			 
Similarly, we extended the model comparison against varying $r,p$ horizons and the SMAPE results are showcased in \cref{subfig:SMAPE_All_methods_best_r_for_p5} in which each curve represents the aggregated SMAPE values across stations $26A-03A$ obtained using best $r$ for each $p$ variation. Once again, BKTR-P and INTR-P seem to outperform the other DL models for short-term predictions $p=1$ and $p=2$ (3min/6min into the future) keeping good performance until $p=3$ (9min ahead). However, after 9min in the future, the performance of the BKTR-P and INTR-P seem to decrease in favour of DL models. This is mostly related to the fact that various DL models applies convolutional layers using multiple historical datasets, whereas the graph-models are bounded by time and space restrictions. However, their performance for short-term predictions is significantly better than for DL models.

\section{CONCLUSIONS}\label{S6_Conclusions}

This paper presents two new graph modelling approaches for motorway flow prediction by using a backtracking and interpolation modelling. Findings reveal that the backtracking algorithm outperforms for short term predictions (less than 10 minutes) all other models, including daily profile prediction, interpolation model and deep learning models (LSTM, CNN, and hybrid CNN-LSTM). This is a proof that simplified graph-modelling approaches only relying on the network structure in both time and space can overcome more complex deep learning approaches for short term predictions; However, more extensive studies need to be conducted for long-term prediction impact of using such methods.  

\textbf{Limitations and future work:}
(1) The algorithms assume no branching structure of the motorway, i.e. we only have one main flow, plus entries and exits. If we had two or more motorways, the traffic at a given station, at a given time point could have originated from multiple points in the past, from all motorways; (2) the methods need to be tested against more complex network structures such as regular urban traffic networks, when the complexity of the graph increases; and also (3) testing the performance of the current models against GCNNs, a popular version of CNN which embeds the graph structure of the network as well (see \citep{YAN2019259}).

{ \small
\section*{ACKNOWLEDGMENT}
This work has been done as part of the ARC Linkage Project LP180100114. The authors are highly grateful for the support of Transport for NSW, Australia. 
}

{ \small
\bibliographystyle{IEEEtranN}

\begin{thebibliography}{18}
\providecommand{\natexlab}[1]{#1}
\providecommand{\url}[1]{#1}
\csname url@samestyle\endcsname
\providecommand{\newblock}{\relax}
\providecommand{\bibinfo}[2]{#2}
\providecommand{\BIBentrySTDinterwordspacing}{\spaceskip=0pt\relax}
\providecommand{\BIBentryALTinterwordstretchfactor}{4}
\providecommand{\BIBentryALTinterwordspacing}{\spaceskip=\fontdimen2\font plus
\BIBentryALTinterwordstretchfactor\fontdimen3\font minus
  \fontdimen4\font\relax}
\providecommand{\BIBforeignlanguage}[2]{{\expandafter\ifx\csname l@#1\endcsname\relax
\typeout{** WARNING: IEEEtranN.bst: No hyphenation pattern has been}\typeout{** loaded for the language `#1'. Using the pattern for}\typeout{** the default language instead.}\else
\language=\csname l@#1\endcsname
\fi
#2}}
\providecommand{\BIBdecl}{\relax}
\BIBdecl

\bibitem[online supplement(2020)]{appendix}
online supplement, ``Appendix: \titlename,'' 2020,
  \url{https://arxiv.org/pdf/2006.14824.pdf#page=9}.

\bibitem[Voort et~al.(1996)Voort, Dougherty, and Watson]{VANDERVOORT1996307}
M.~V.~D. Voort, M.~Dougherty, and S.~Watson, ``Combining kohonen maps with
  arima time series models to forecast traffic flow,'' \emph{Transportation
  Research Part C}, vol.~4, no.~5, pp. 307 -- 318, 1996.

\bibitem[Khoei et~al.(2013)Khoei, Bhaskar, and Chung]{Khoei2013}
A.~M. Khoei, A.~Bhaskar, and E.~Chung, ``Travel time prediction on signalised
  urban arterials by applying sarima modelling on bluetooth data,'' in
  \emph{Australasian Transport Research Forum}, 2013.

\bibitem[Williams(2001)]{Williams2001}
B.~M. Williams, ``Multivariate vehicular traffic flow prediction: Evaluation of
  arimax modeling,'' \emph{Trans. Res. Rec.}, vol. 1776, no.~1, pp. 194--200,
  2001.

\bibitem[Dougherty and Cobbett(1997)]{DOUGHERTY199721}
M.~S. Dougherty and M.~R. Cobbett, ``Short-term inter-urban traffic forecasts
  using neural networks,'' \emph{International Journal of Forecasting},
  vol.~13, no.~1, pp. 21 -- 31, 1997.

\bibitem[Lv et~al.(2015)Lv, Duan, Kang, Li, and Wang]{Lv2015}
Y.~Lv, Y.~Duan, W.~Kang, Z.~Li, and F.~Y. Wang, ``{Traffic Flow Prediction with
  Big Data: A Deep Learning Approach},'' \emph{IEEE Transactions on Intelligent
  Transportation Systems}, vol.~16, no.~2, pp. 865--873, 2015.

\bibitem[Zhang et~al.(2016)Zhang, Zheng, and Qi]{Zhang2016}
J.~Zhang, Y.~Zheng, and D.~Qi, ``Deep spatio-temporal residual networks for
  citywide crowd flows prediction,'' \emph{CoRR}, vol. abs/1610.00081, 2016.

\bibitem[Polson and Sokolov(2017)]{POLSON20171}
N.~G. Polson and V.~O. Sokolov, ``Deep learning for short-term traffic flow
  prediction,'' \emph{Trans. Research Part C}, vol.~79, pp. 1 -- 17, 2017.

\bibitem[Dai et~al.(2017)Dai, Fu, Lin, Li, and Wang]{Dai2017}
X.~Dai, R.~Fu, Y.~Lin, L.~Li, and F.~Wang, ``Deeptrend: {A} deep hierarchical
  neural network for traffic flow prediction,'' \emph{CoRR}, vol.
  abs/1707.03213, 2017.

\bibitem[{Jia} et~al.(2017){Jia}, {Wu}, {Ben-Akiva}, {Seshadri}, and
  {Du}]{Jia2017}
Y.~{Jia}, J.~{Wu}, M.~{Ben-Akiva}, R.~{Seshadri}, and Y.~{Du},
  ``Rainfall-integrated traffic speed prediction using deep learning method,''
  \emph{IET Intelligent Transport Systems}, vol.~11, no.~9, pp. 531--536, 2017.

\bibitem[Wu et~al.(2018)Wu, Tan, Qin, Ran, and Jiang]{WU2018166}
Y.~Wu, H.~Tan, L.~Qin, B.~Ran, and Z.~Jiang, ``A hybrid deep learning based
  traffic flow prediction method and its understanding,'' \emph{Transportation
  Research Part C: Emerging Technologies}, vol.~90, pp. 166 -- 180, 2018.

\bibitem[{Mihaita} et~al.(2019){Mihaita}, {Li}, {He}, and
  {Rizoiu}]{Mihaita2019}
A.~{Mihaita}, H.~{Li}, Z.~{He}, and M.~{Rizoiu}, ``Motorway traffic flow
  prediction using advanced deep learning,'' in \emph{2019 IEEE Intelligent
  Transportation Systems Conference (ITSC)}, Oct 2019, pp. 1683--1690.

\bibitem[Wang et~al.(2018)Wang, Chen, Min, He, Yang, and Zhang]{Wang2018}
X.~Wang, C.~Chen, Y.~Min, J.~He, B.~Yang, and Y.~Zhang, ``Efficient
  metropolitan traffic prediction based on graph recurrent neural network,''
  \emph{CoRR}, vol. abs/1811.00740, 2018.

\bibitem[Yu et~al.(2017)Yu, Yin, and Zhu]{Yu2017}
B.~Yu, H.~Yin, and Z.~Zhu, ``Spatio-temporal graph convolutional neural
  network: {A} deep learning framework for traffic forecasting,'' \emph{CoRR},
  vol. abs/1709.04875, 2017.

\bibitem[Fang et~al.(2019)Fang, Zhang, Meng, Xiang, and Pan]{Shen2019}
S.~Fang, Q.~Zhang, G.~Meng, S.~Xiang, and C.~Pan, ``Gstnet: Global
  spatial-temporal network for traffic flow prediction,'' in
  \emph{International Joint Conference on Artificial Intelligence,
  {IJCAI-19}}.\hskip 1em plus 0.5em minus 0.4em\relax International Joint
  Conferences on Artificial Intelligence Organization, 7 2019, pp. 2286--2293.

\bibitem[{Mihaita} et~al.(2020){Mihaita}, {Li}, and {Rizoiu}]{Mihaita2020}
\BIBentryALTinterwordspacing
A.~{Mihaita}, H.~{Li}, and M.~{Rizoiu}, ``Traffic congestion anomaly detection
  and prediction using deep learning,'' \emph{arXiv draft arXiv:2006.13215},
  2020. [Online]. Available: \url{https://arxiv.org/abs/2006.13215}
\BIBentrySTDinterwordspacing

\bibitem[Paszke et~al.(2017)Paszke, Gross, Chintala, Chanan, Yang, DeVito, Lin,
  Desmaison, Antiga, and Lerer]{paszke2017automatic}
A.~Paszke, S.~Gross, S.~Chintala, G.~Chanan, E.~Yang, Z.~DeVito, Z.~Lin,
  A.~Desmaison, L.~Antiga, and A.~Lerer, ``Automatic differentiation in
  pytorch,'' in \emph{NIPS-W}, 2017.

\bibitem[Yan et~al.(2019)Yan, Ai, Yang, and Yin]{YAN2019259}
X.~Yan, T.~Ai, M.~Yang, and H.~Yin, ``A graph convolutional neural network for
  classification of building patterns using spatial vector data,'' \emph{ISPRS
  Journal of Photogrammetry and Remote Sensing}, vol. 150, pp. 259 -- 273,
  2019.

\end{thebibliography}

}

\clearpage
 \appendix

This document is accompanying the submission \textit{\titlename}.
 The information in this document complements the submission, and it is presented here for completeness reasons. It is not required for understanding the main paper, nor for reproducing the results.
\begin{algorithm}
	\caption{
		\textbf{INTR-P:} predict traffic flow using interpolation, for a single station on one day.
	}
	\begin{algorithmic} 
	\REQUIRE $G(V, E)$ -- graph of stations along motorway;\\
			 $V$ -- set of stations; $E$ -- set of geographical relations between stations\\
			 $p$ -- number of steps in the future to predict flow;\\
			 $r$ -- number of time intervals in the past to use;\\
			 $v$ -- station for which to predict the flow;\\
			 $F(\cdot; \{t-r, t-r+1, .. , t-1\})$ -- observed flow for all stations from $t-p$ until present;\\
			 $\overline{F}(\cdot; \{t+1, .. , t+p-1\})$ -- historical means for all stations at future time points.
	\ENSURE $\widehat{F}(v ; t+p)$ -- traffic flow for station $v$ at $p$ intervals in the future.
	\STATE \textit{\textcolor{gray}{// find optimal $u$ for target $v$;}}
	\STATE $optim\_dist \leftarrow (r+p) \times d \times speed$ ;
	\STATE Find $u \in V$ of type $a$ so that 
	$len(u \rightarrow s_k) + \sum_{j=k}^{l-1} len(s_j \rightarrow s_{j+1}) + len(s_l \rightarrow v) \approx optim\_dist,$ with $s_k, .., s_l \in V$ ;
	\STATE
	\STATE \textit{\textcolor{gray}{// traverse from $u$ to $v$ to identify entries and exits;}}
	\STATE $\mathcal{E} \leftarrow \emptyset$; $\mathcal{X} \leftarrow \emptyset$;
	\STATE $\mathcal{B} \leftarrow \{u\}$; \textit{\textcolor{gray}{// Breadth First Search node set}}
	\STATE $dist\_from\_u \leftarrow 0$;
	\WHILE{$\mathcal{B} \neq \emptyset$}
		\STATE pop $head$ from $\mathcal{B}$;  
		\IF{$head$ of type $a$}
			\STATE \textit{\textcolor{gray}{// update the distance from $u$}}
			\STATE $dist\_from\_u \leftarrow dist\_from\_u + len(head, p(head))$
			\STATE \textit{\textcolor{gray}{// add stations linked to $head$ to the queue}}
			\STATE \textit{\textcolor{gray}{// first those of type $a$, then entries and exits}}
			\STATE $\mathcal{B} \leftarrow \mathcal{B} \cup \{s\}$, $\forall s \in V$ with $head \rightarrow s, type(s) = a$
			\STATE $\mathcal{B} \leftarrow \mathcal{B} \cup \{s\}$, $\forall s \in V$ with $head \rightarrow s, type(s) \in \{e, x\}$
			\STATE $p(s) \leftarrow head, \forall s \in V$ with $head \rightarrow s$
		\ENDIF
		\IF{$head$ of type $e$}
			\STATE \textit{\textcolor{gray}{// add $head$ to entries group}}
			\STATE $\mathcal{E} \leftarrow \mathcal{E} \cup \{head\}$;
		\ENDIF
		\IF{$head$ of type $x$}
			\STATE \textit{\textcolor{gray}{// add $head$ to exits group}}
			\STATE $\mathcal{X} \leftarrow \mathcal{X} \cup \{head\}$;
		\ENDIF
		\STATE compute $align(head)$ on time intervals based on $dist\_from\_u$;
	\ENDWHILE	
	
	\STATE
	\STATE \textit{\textcolor{gray}{// compute predicted flow for $v$;}}
	\STATE $\widehat{F}(v ; t+p) \leftarrow F(u; t-p)$
\STATE \textit{\textcolor{gray}{// add traffic of entries, subtract traffic of exits}}
	\FOR{each $s \in \mathcal{E} \bigcup \mathcal{X}$}
		\STATE $i \leftarrow align(s)$
		\IF{$i+1 < p$}
			\STATE $F^*(s; i \rightarrow i+1) = \dfrac{t_1 F(s; i)}{d} +\dfrac{t_2 F(s; i+1)}{d}$
		\ELSE
			\STATE $F^*(s; i \rightarrow i+1) = \dfrac{t_1 F(s; i)}{d} +\dfrac{t_2 \overline{F}(s; i+1)}{d}$
		\ENDIF
		
		\IF{$s \in \mathcal{E}$}		
			\STATE $\widehat{F}(v ; t+p) \leftarrow \widehat{F}(v ; t+p) + F^*(e; i \rightarrow i+1)$ \ELSE
			\STATE $\widehat{F}(v ; t+p) \leftarrow \widehat{F}(v ; t+p) - F^*(e; i \rightarrow i+1)$ \ENDIF
	\ENDFOR

	\RETURN $\widehat{F}(v ; t+p)$
	\end{algorithmic}
	\label{si-algo:interpolation}
\end{algorithm} 
\section*{DPP metrics before and after outlier detection and anomaly cleaning}

\begin{figure*}[tbp]
	\centering
	\newcommand\myheight{0.205} 

	\subfloat[]{
		\includegraphics[width=0.49\textwidth]{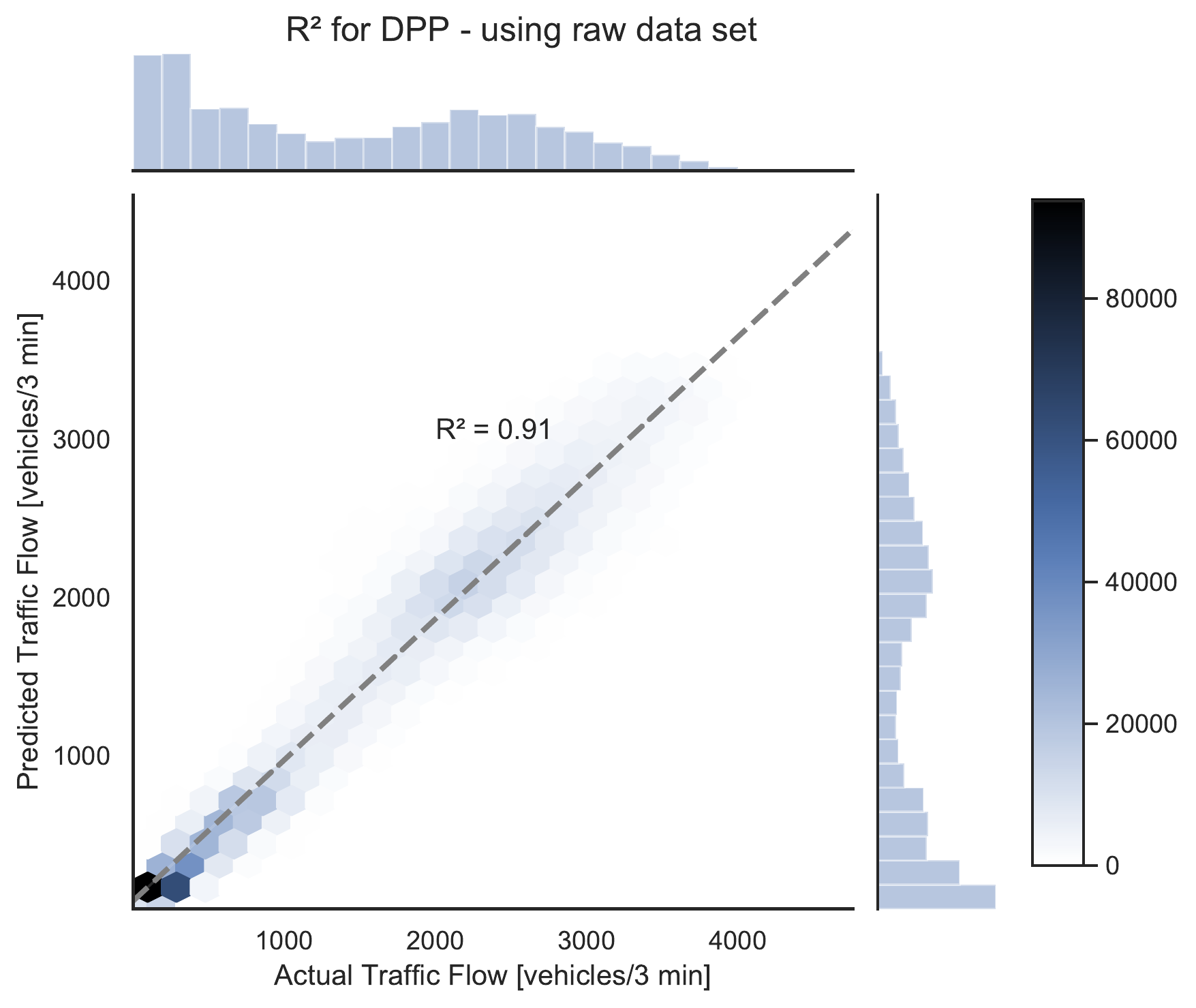}\label{subfig:DPP_R2_r1p1_raw}}
	\subfloat[]{
	    \includegraphics[width=0.49\textwidth]{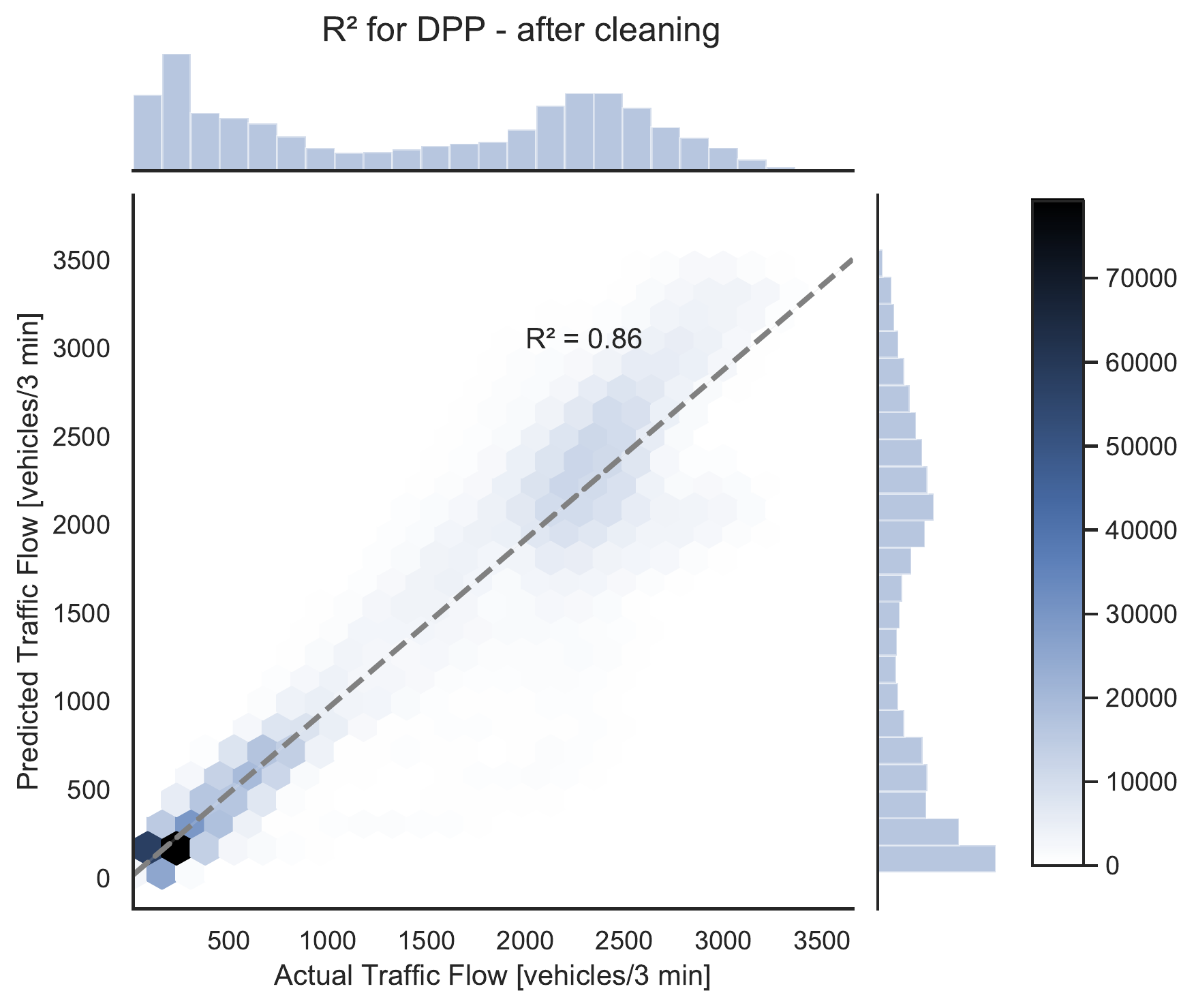}\label{subfig:DPP_R2_r1p1_clean}}
	\\
	\subfloat[]{
		\includegraphics[width=0.49\textwidth]{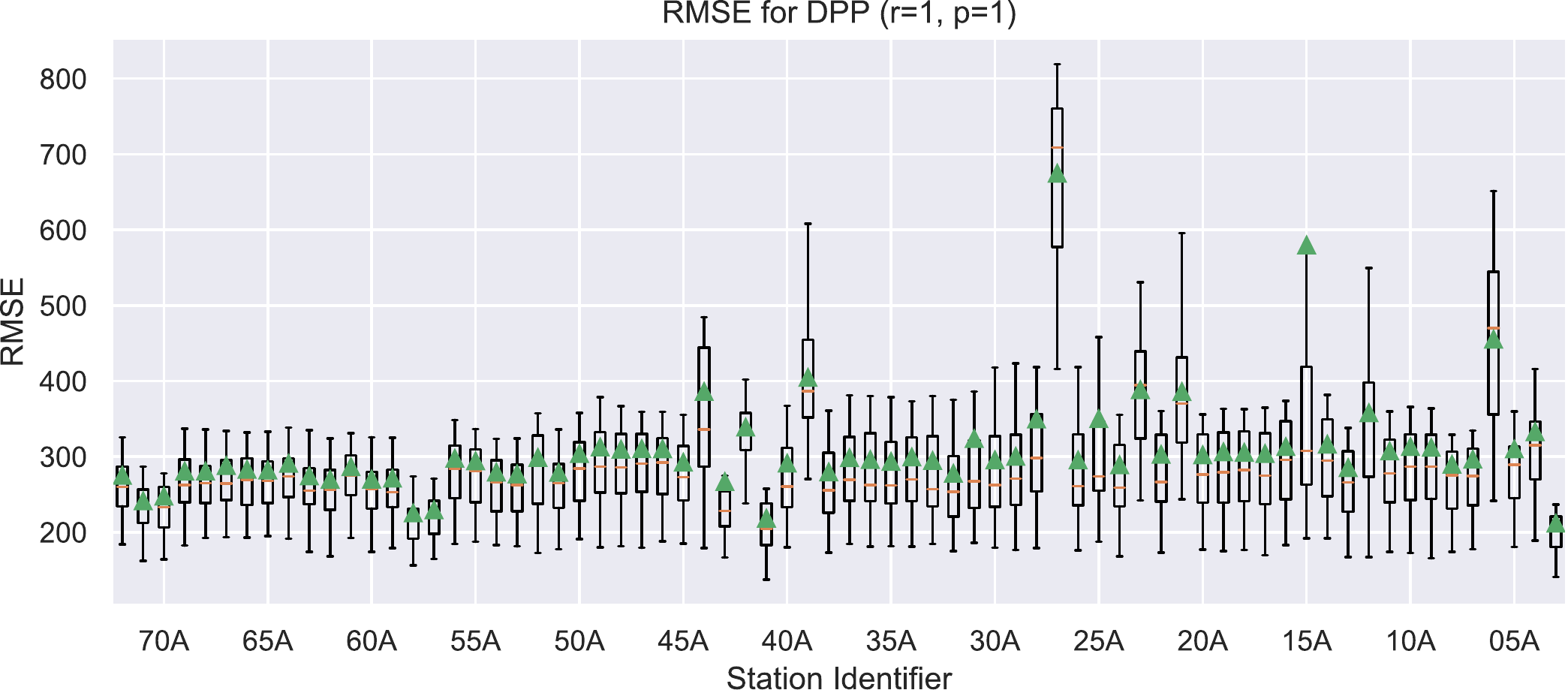}\label{subfig:DPP_RMSE_r1p1_raw}}
	\subfloat[]{
	    \includegraphics[width=0.49\textwidth]{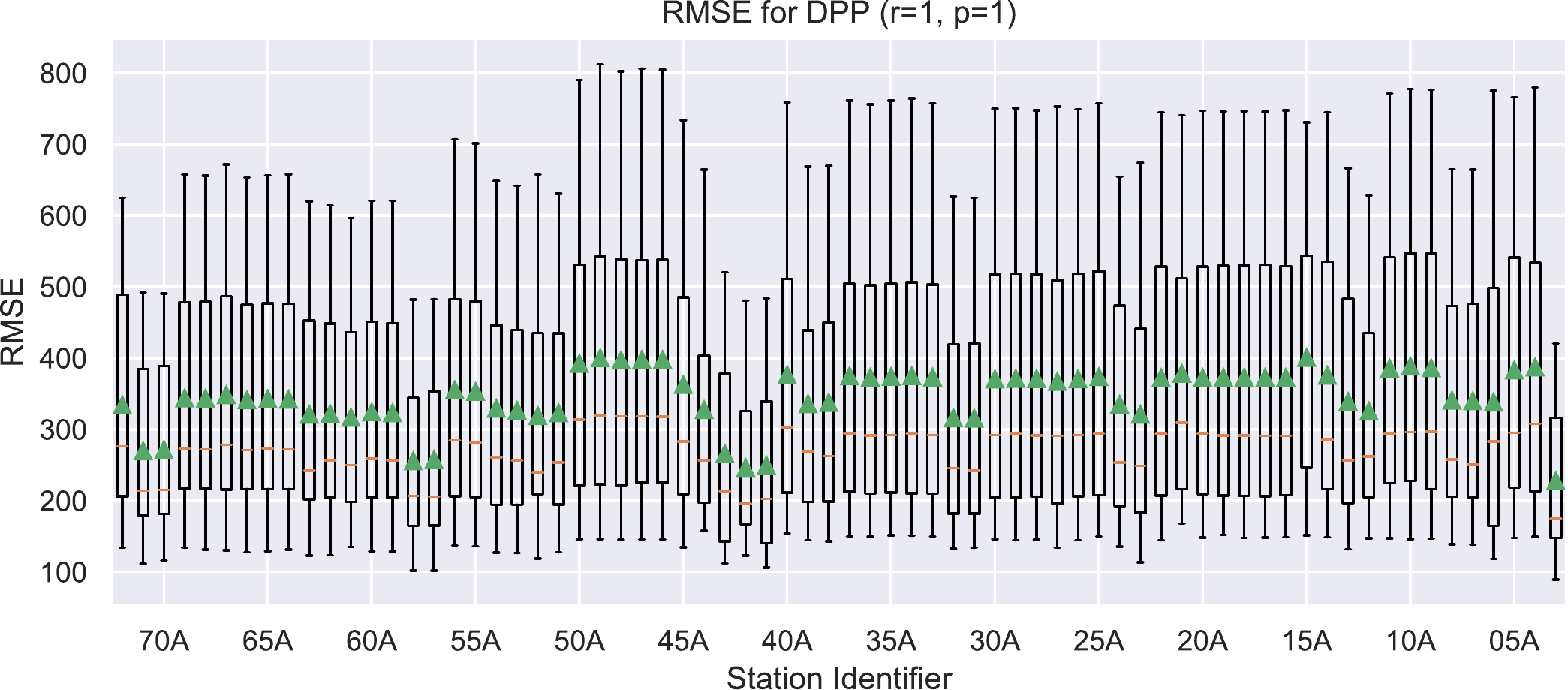}\label{subfig:DPP_RMSE_r1p1_clean}}
	\\
	
	\subfloat[]{
		\includegraphics[width=0.49\textwidth]{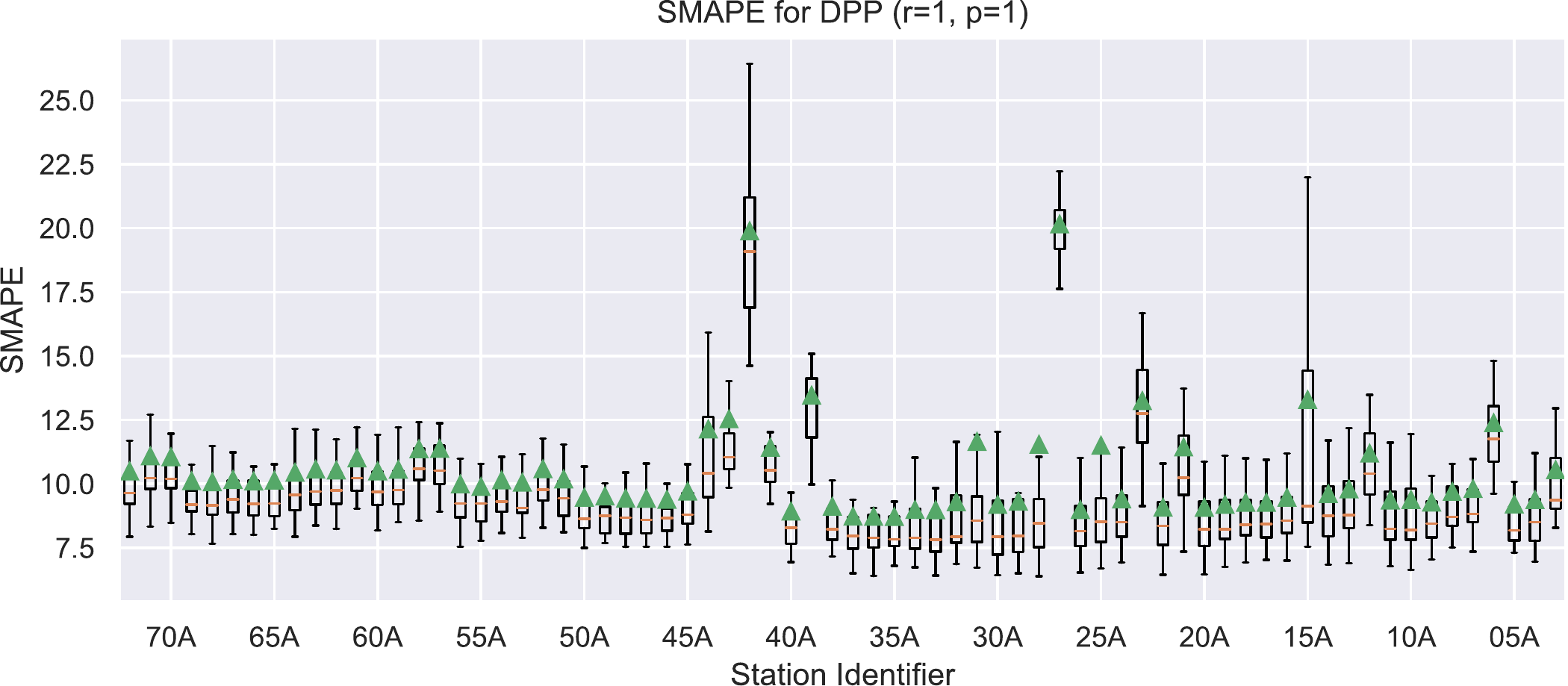}\label{subfig:DPP_SMAPE_r1p1_raw}}
	\subfloat[]{
	    \includegraphics[width=0.49\textwidth]{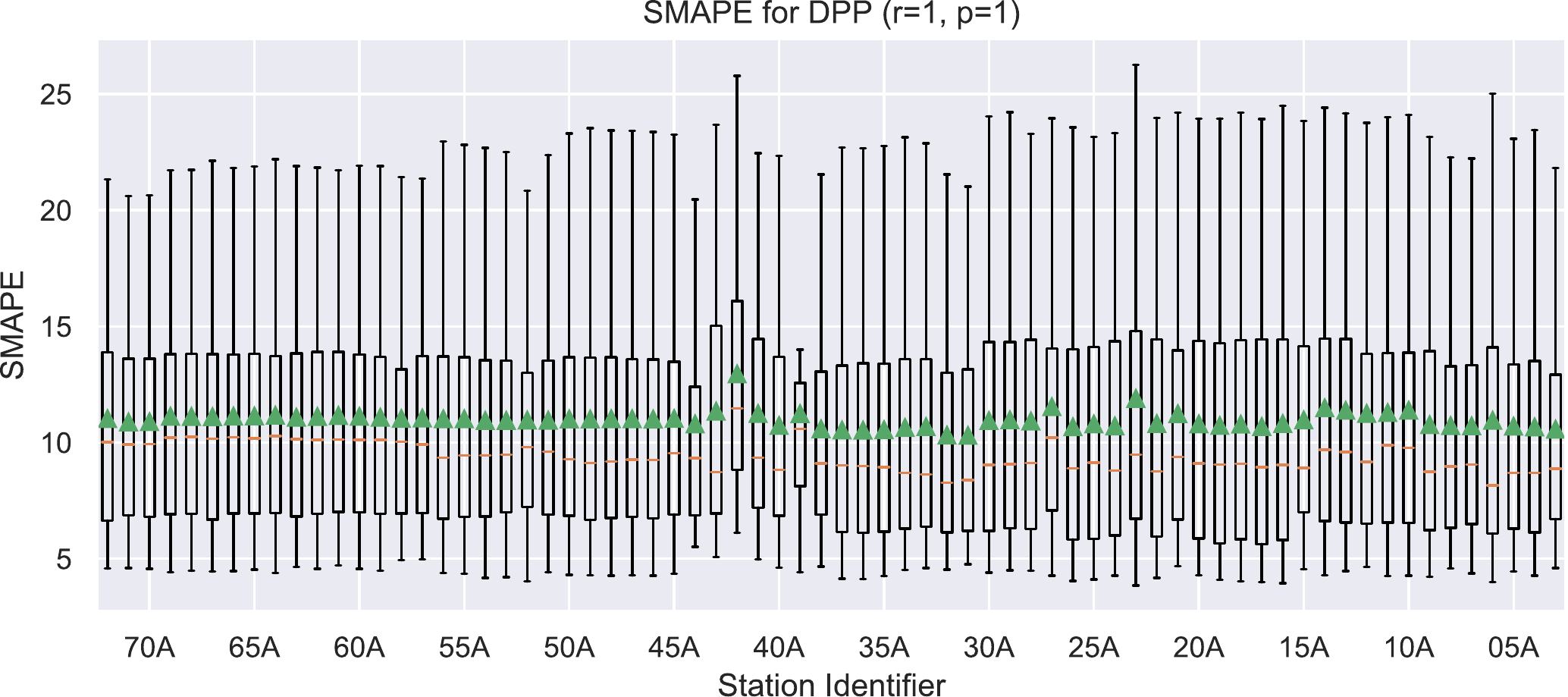}\label{subfig:DPP_SMAPE_r1p1_clean}}
	\caption{DPP metrics before and after data cleaning: 
	a) $R^2$ for DPP using raw data set, 
	b) $R^2$ for DPP after cleaning,
	c) RMSE for DPP using raw data set, 
	d) RMSE for DPP after cleaning,
	e) SMAPE for DPP using raw data set, 
	f) SMAPE for DPP after cleaning.
	}
	\label{fig:DPP_before_after_cleaning}
\end{figure*}

\cref{fig:DPP_before_after_cleaning} represents the performance metrics ($R^2$, RMSE and SMAPE) of the Daily Profile Predictor obtained before and after the outlier detection and anomaly cleaning algorithms have been applied. The comparison showcases a considerable improvement across all metrics, which previously were affected by severe outliers (missing data or very large vehicles quantities). This was the first step in identifying problems during the data profiling procedure and helped in addressing some of the challenges of building prediction methods using graph-based approaches as severe outliers of a neighbourhood station would highly impact performance of prediction of another station. \cref{fig:DPP_before_after_cleaning}a) and b) showcase more specifically the density plot obtained from the $R^2$ value of DPP before/after data cleaning and reflects two major peaks in terms of density points: one for traffic flows less than $1000 veh/T_i$ and another one for traffic flow between $2200-2500 veh/T_i$, mostly explained of the two-peak daily data profiles recorded for each station.  

\begin{figure}[tbp]
	\centering
	\newcommand\myheight{0.205} 

	\subfloat[]{
		\includegraphics[width=0.49\textwidth]{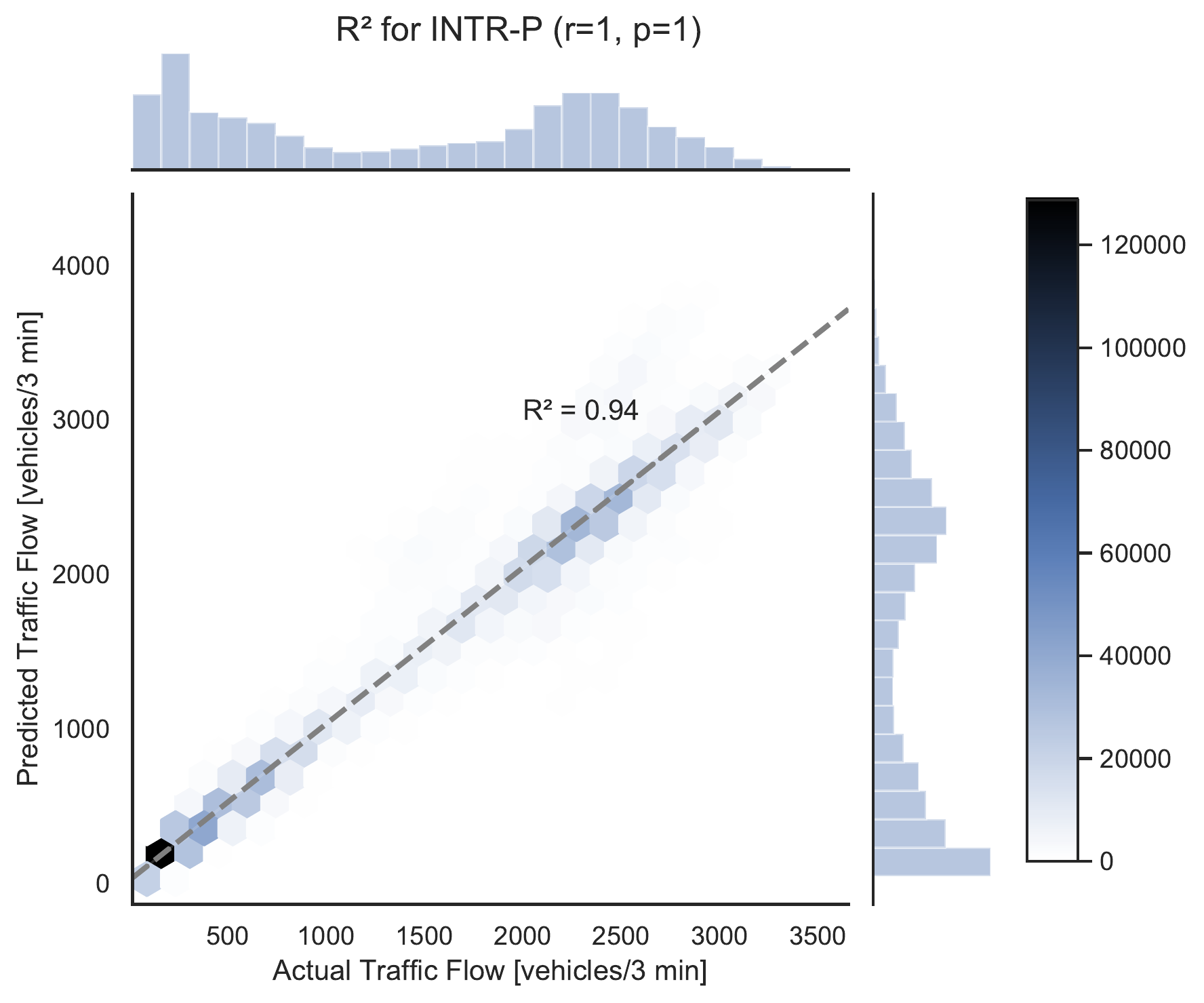}\label{subfig:INTRP_R2_r1p1}}
	\\
	\subfloat[]{
	    \includegraphics[width=0.49\textwidth]{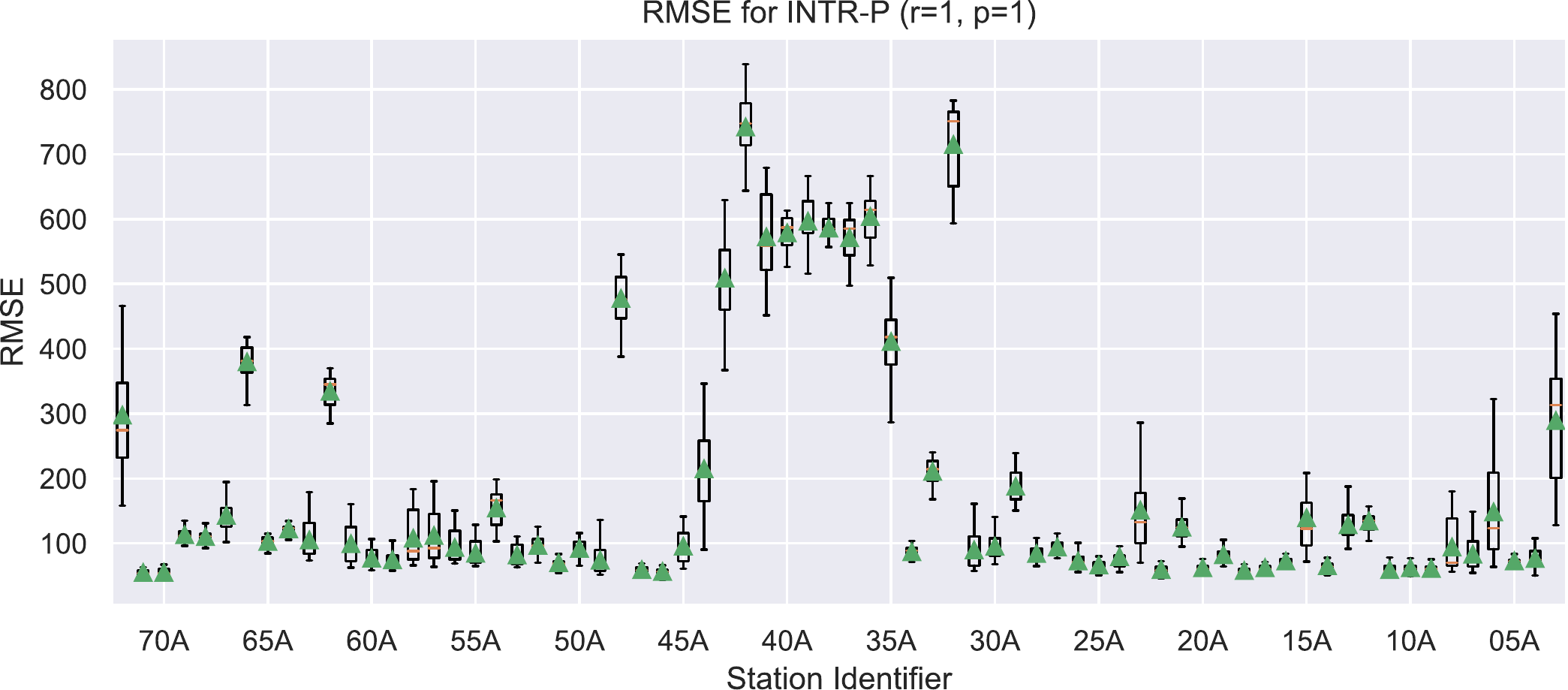}\label{subfig:INTRP_RMSE_r1p1}}
	\\
	\subfloat[]{
		\includegraphics[width=0.49\textwidth]{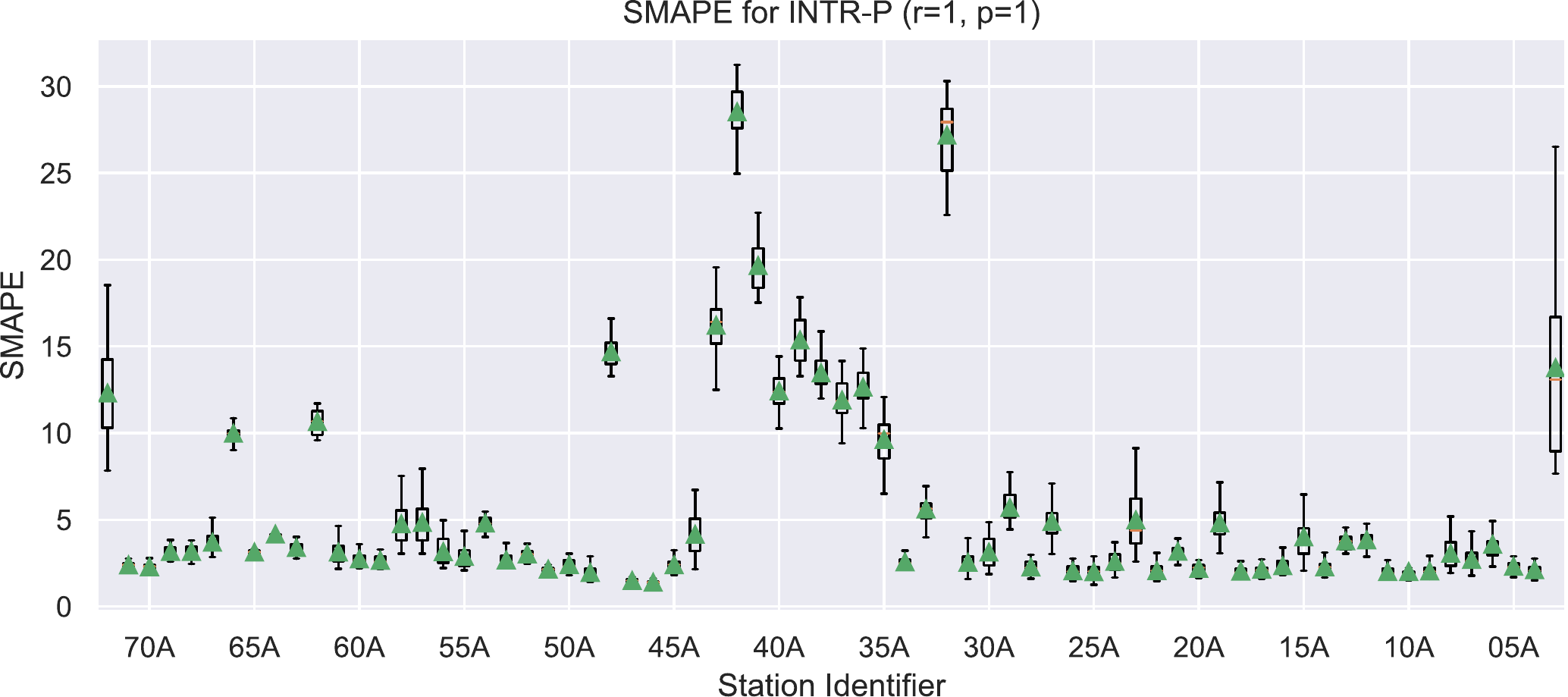}\label{subfig:INTRP_SMAPE_r1p1}}
		
	\caption{INTR-P metrics: a) $R^2$, b) RMSE, c) SMAPE.} 
		
	\label{fig:INTR_metrics}
\end{figure}

\begin{figure}[tbp]
	\centering
	\newcommand\myheight{0.205} 

	\subfloat[]{
	    \includegraphics[width=0.4\textwidth]{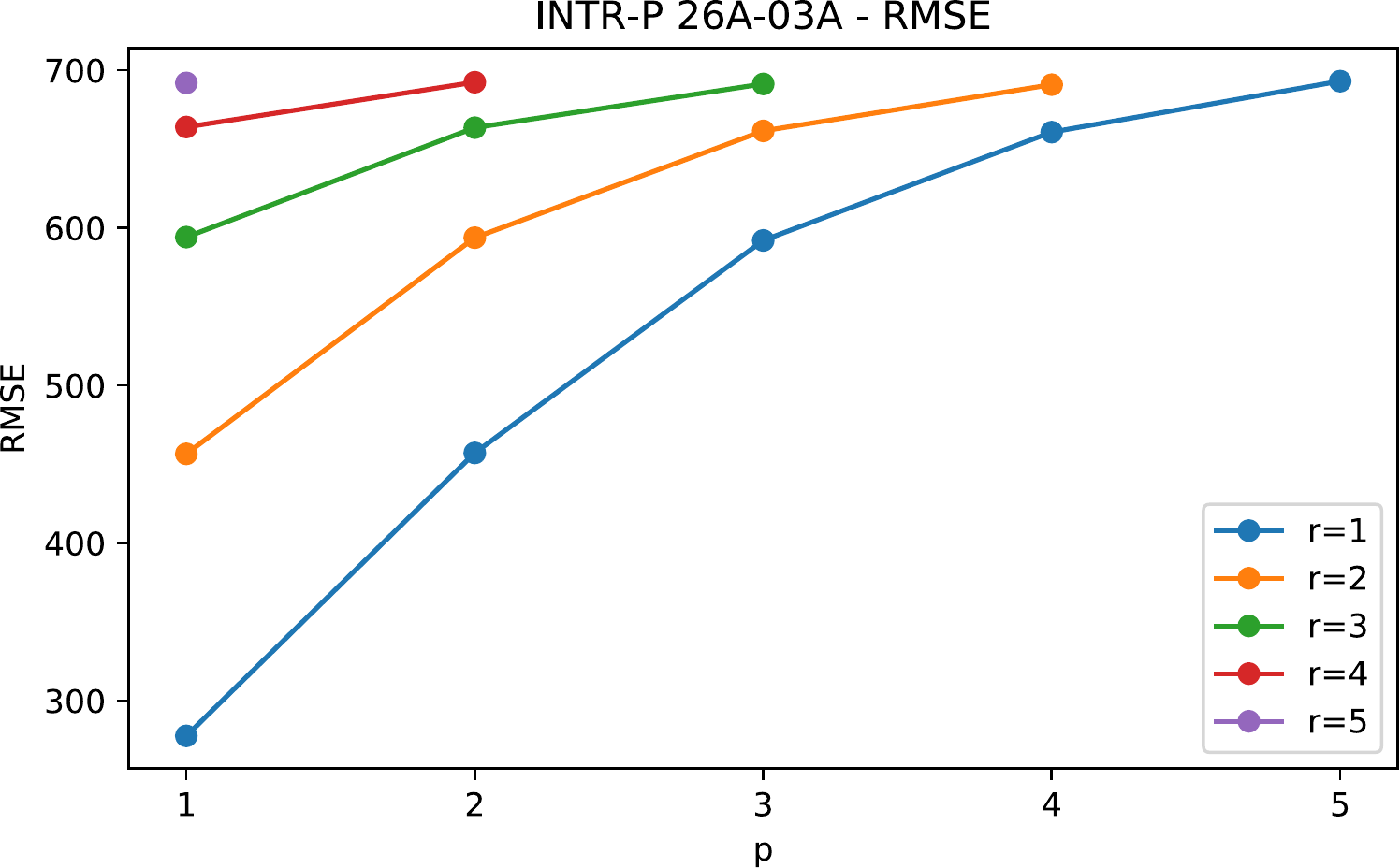}\label{subfig:INTRP_RMSE_r1p5}}
	\\
	\subfloat[]{
		\includegraphics[width=0.4\textwidth]{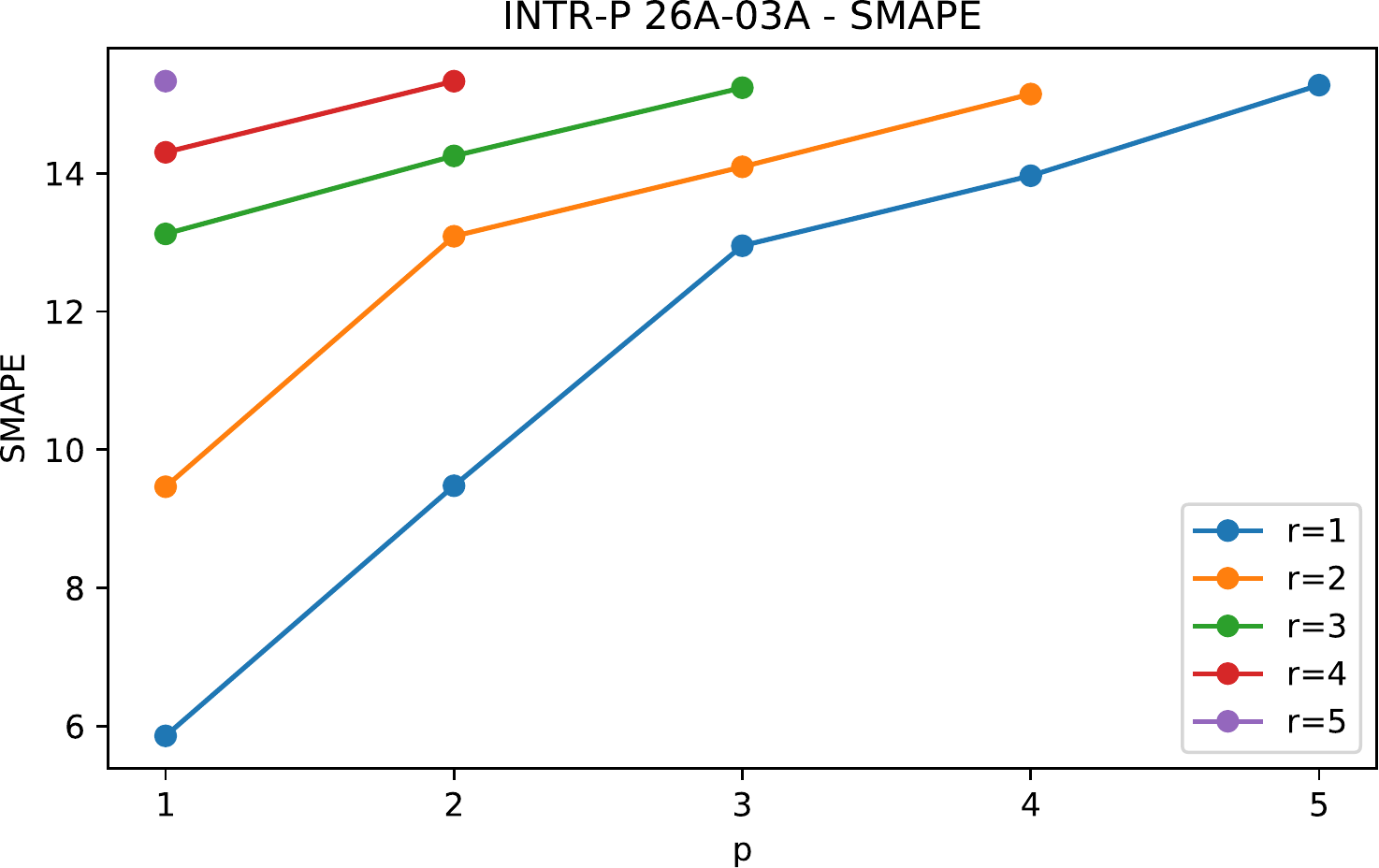}\label{subfig:INTRP_SMAPE_r1p5}}
		
	\caption{a) INTR-P results for various $r$, $p$: a)RMSE b) SMAPE.} 
		
	\label{fig:INTRP_P_Past_future_horizons}
\end{figure}

\begin{figure}[tbp]
	\centering
	\newcommand\myheight{0.205} 

	\subfloat[]{
		\includegraphics[width=0.4\textwidth]{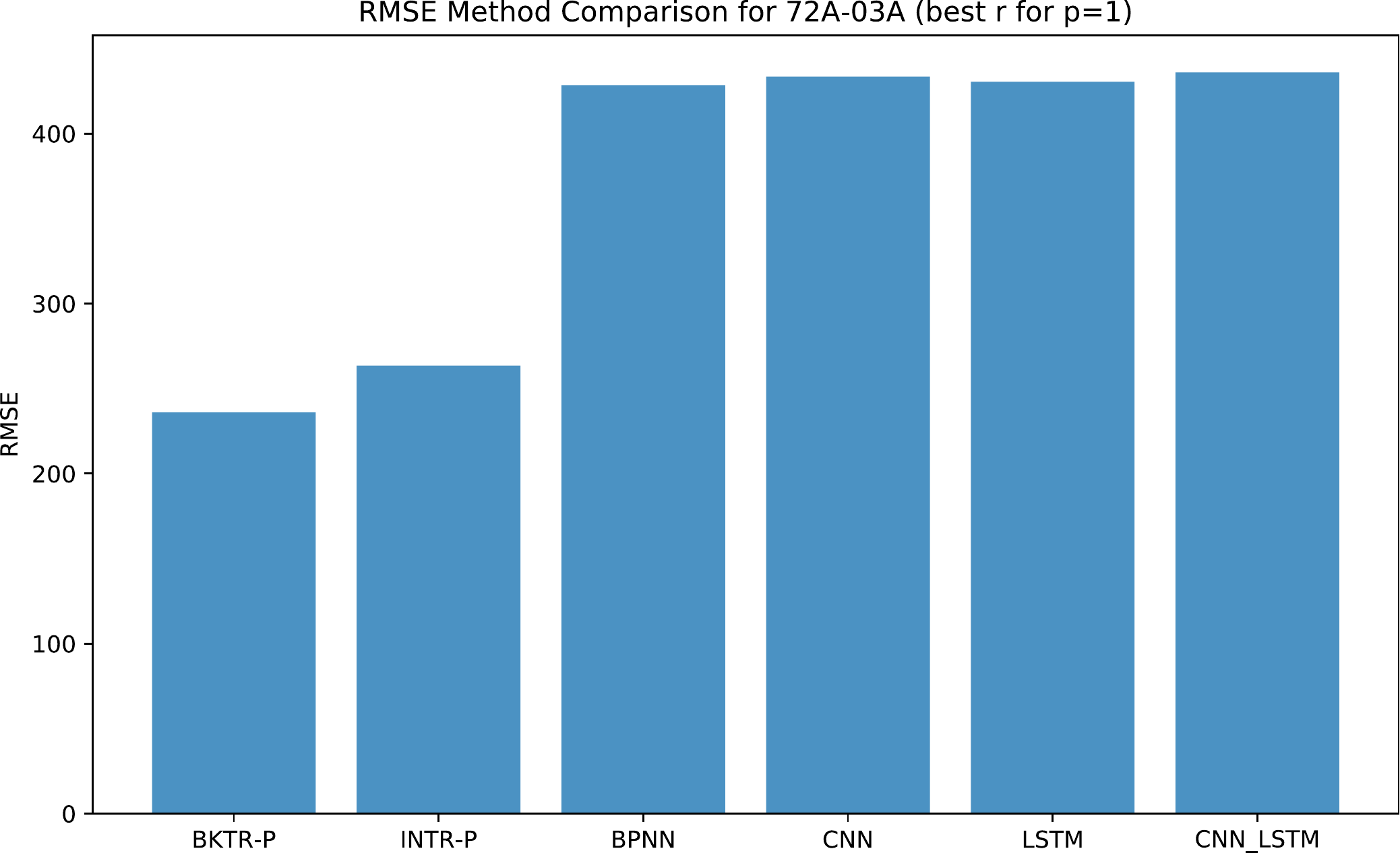}\label{subfig:RMSE_All_methods_best_r_for_p1}}	
	\\
	\subfloat[]{
		\includegraphics[width=0.4\textwidth]{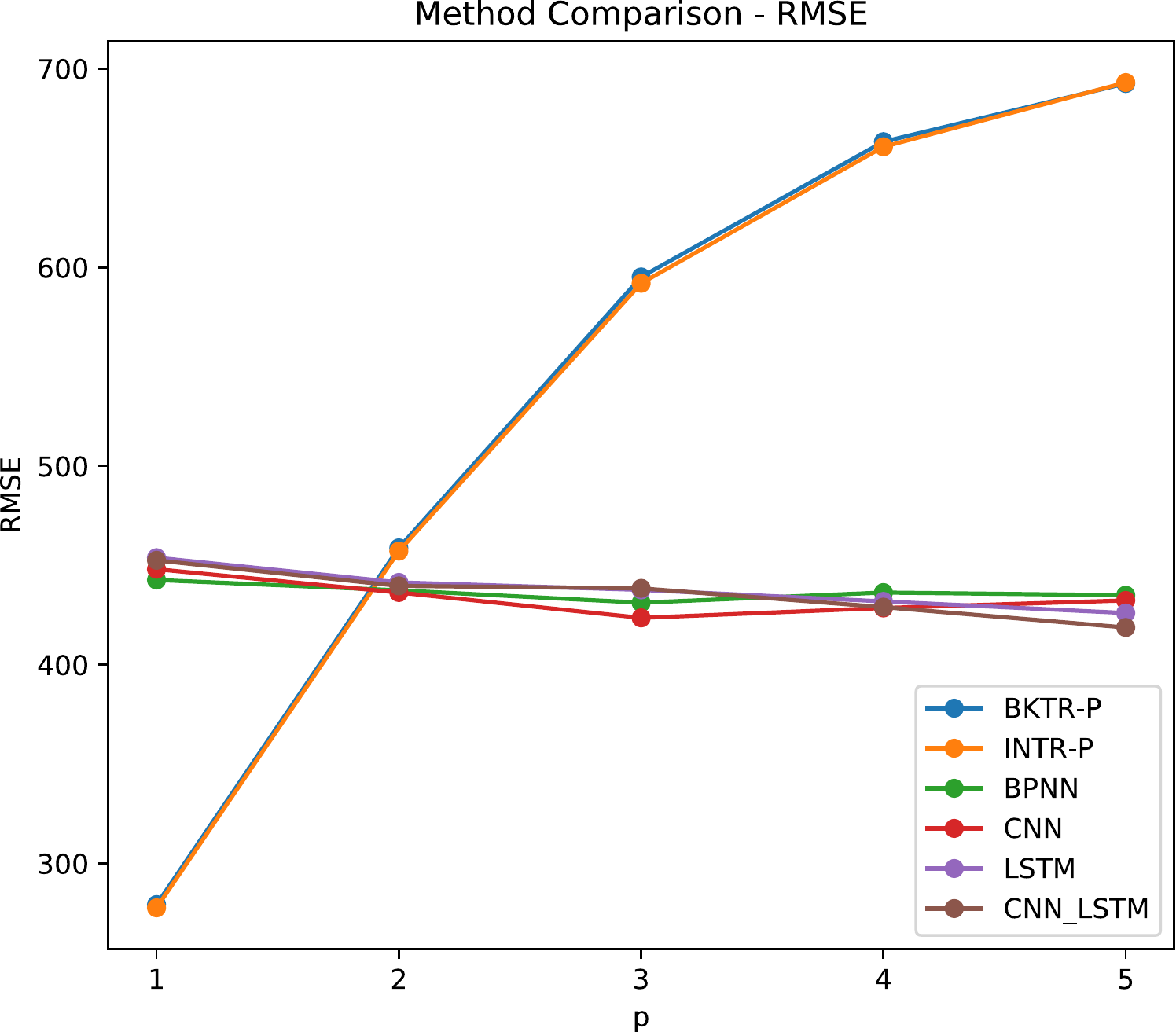}\label{subfig:RMSE_All_methods_best_r_for_p5}}
			
	\caption{ All methods comparison a) RMSE using best r for p=1, b) RMSE using best r for $p\in\{1,..5\}$.} 
		
	\label{fig:RMSE_Past_future_horizons_best_p5}
\end{figure}

\end{document}